\documentclass[preprint,3p,number]{elsarticle}
\usepackage[T1]{fontenc}
\usepackage[utf8]{inputenc} %Polish input (dedication at the end!)
\usepackage{framed} %
\usepackage{amsmath}
\usepackage{amsfonts}
\usepackage{amssymb}
\usepackage{latexsym}
\usepackage{graphics}
\usepackage{graphicx}
\usepackage{epigraph}
\usepackage{framed} 
\usepackage{epsf}
\usepackage{aniszewski,color}
\title{Improvements,  testing and development of the {ADM}-{$\tau$} sub-grid surface tension model for two-phase {LES}}
\author[coria]{Wojciech Aniszewski\corref{cor1}}
\ead{aniszewski@dalembert.upmc.fr}

\cortext[cor1]{Corresponding Author at:  Institut Jean Le Rond d'Alembert, Universit\'e Pierre et Marie Curie, 4 Place Jussieu, 75252 Paris, France. tel: (+33)144-278-714, email: aniszewski@dalembert.upmc.fr, aniszewski@protonmail.ch}

\address[dalembert]{Sorbonne Universit\'es, UPMC Univ Paris 06, CNRS UMR 7190 - Institut Jean Le Rond d'Alembert, Paris, France}

\journal{Journal of Computational Physics}

\begin{document}
\bibliographystyle{plain}
%\linenumbers

\begin{abstract}
  In this paper, a specific subgrid term occurring in Large Eddy Simulation (LES) of two-phase flows is investigated. This and other subgrid terms are presented, we subsequently elaborate on the existing models for those and re-formulate the ADM$-\tau$ model for sub-grid surface tension previously published by these authors. This paper presents a substantial, conceptual  simplification over the original model version, accompanied by a decrease in its computational cost. At the same time, it addresses the issues the original model version faced, e.g. introduces non-isotropic applicability criteria based on resolved interface's principal curvature radii. Additionally, this paper introduces more throughout testing of the ADM-$\tau,$ in both simple and complex flows.
\end{abstract}


\maketitle

%% \begin{framed}
%%   \tableofcontents
%% \end{framed}

\section{Introduction}

The defining characteristic of most two-phase flows described in Eulerian manner is the presence of the interface. Both physics of the flow and its mathematical description complicate once one settles upon incorporating this \textit{surface of discontinuity} which not only influences the balance of forces in the fluids, but itself continuously evolves, changing its topology.  It is also long established that the interfaces may be of multiscale character \cite{cd_part1}. For example, the elusive mechanism of the liquid core atomization is, as believed today, based on the coexistence of the large-scale (Kelvin-Helmholtz) instabilities near the injection core \cite{devillers} and capillary instabilities that may appear downstream in liquid formations orders of magnitude smaller \cite{shinjo}. There are many unknowns not only pertaining the distribution of scales in such system, but also concerning the manner in which energy conversion is realized during its spatio-temporal evolution. Combine this with the fact that in many two-phase flows turbulence may occur, with kinetic energy transfer from liquid to gas (Diesel injector \cite{devillers}) or \textit{vice versa} (coaxial injectors \cite{geo2015}); and it will appear obvious that immense resolutions are needed to simulate this kind of phenomena with any hopes of capturing their enormous complexity.

Whenever such a resolution is available it is never truly adequate, especially in realistic cases. For example, impressive calculations presented by Shinjo and Umemura \cite{shinjo} still don't amount for a fully-resolved, low-Reynolds number atomization  simulation even if supercomputers were generously employed. Those simulating higher $Re$ regimes must settle for incompleteness as their simulations, while capturing parts of the mechanisms at hand and providing important insights \cite{menard,devillers} remain largely under-resolved. Most of those works treat the small-scale phenomena directly (Direct Numerical Simulation, DNS \cite{menard}) or try to model small-scale (sub-grid) phenomena in single-phase flow specific manner \cite{devillers}.

The \textit{first} of those approaches could be termed ``assumed DNS'', as it takes all the important scales as resolved/represented. Some authors use the term ``implicit LES'' which seems dual with ''assumed DNS'' as both amount to no consideration of small scales whatsoever, except that the  former claims to include them indirectly  by means of  \textit{numerical dissipation} (the apparent energy dissipation caused by discretization). This may be correct in some situations, also it definitely results in robustness and simplicity as no additional assumptions are introduced into the methodology of Navier-Stokes equations solving. However, due to immense range of scales in turbulent flows -- no less two-phase -- the question whether the small-scales are ''correctly'' represented here remains, at best, open.

The \textit{second} approach is to use Large Eddy Simulation (LES) which solves explicitly for large (resolvable\footnote{Those concepts are dual in the 'grid-filter' approach applied in this paper.}) scales, and models the influence of sub-grid (sub-filter) scales \cite{geurts}. It is widespread in single-phase flow modelling and nonetheless in modelling of some two-phase flow cases \cite{devillers, bianchi}. The usage of single-phase LES models (Smagorinsky \cite{smagorinsky}, Structure Functions \cite{pope}) is also very popular within the scope of Lagrangian-type two-phase simulations. Here, using LES models the velocity field is said to have a broadened content of small-scale (high-frequency information injected by --in most approaches -- the mechanism of ''turbulent viscosity'' \cite{pope}. Then, the second phase regions (particles in Lagrangian methods) interact with this field. In case of Eulerian methods, the interaction is advection of the phase-indicator function. As the single-fluid formulation for two-phase flows involves more nonlinear terms than that for single-phase (surface tension, advection equation) the LES-specific filtering operation, if applied to those terms would require more closures than in the single-phase formulation. Since this paper is devoted to the closure of one of these terms, we will now briefly review this issue.

The first systematic study of such terms is due to Alajbegovic \cite{alajbegovic}. In his brief conference paper he has even included a simple model proposed to close the sub-grid surface tension term. This model has however remained untested as the author has discontinued the work. However the formalism presented there is fully correct \cite{liovic2012}. Later accounts of the two-phase specific terms due to LES filtering operation are mainly due to the group of St\'ephane Vincent \cite{vincent, toutant, labourasse} who have provided rigorous formal filtration of all two-phase flow specific terms, as well as \textit{a priori} calculations (using explicit filtering of DNS results) of some of them. These works have showed that in specific flow configurations (such as the phase inversion problem, see Section \ref{sephase_sect}) the interface-specific sub-grid contributions can not be neglected, as a significant input e.g. into the momentum balance would be left out. This paper shares most of its nomenclature with Vincent's work.

%Among others, the sub-grid surface tension is represented here as the $\trnn$ tensor, so following \cite{vincent}. LES-specific filtering operation being linear, this term stems from the fact that the surface tension force includes a product of curvature and normal vector, both of which are filtered\footnote{For mathematical accuracy, it should be mentioned that also the Dirac $\delta$ distribution, restricting the term to the interface is filtered.}; thus a closure is required much like for the momentum term. In fact, the early closure proposed by Alajbegovic was very similar to the residual stress tensor model of Smagorinsky in that it also related $\trnn$ to the filtered rate of strain (and, as noted by Liovic \& Lakehal \cite{liovic2012}, without much justification).

To this authors' knowledge, at least three possible closures have been published -- roughly at the beginning of the decade -- for the sub-grid surface tension term. Due to overlap in publication cycles all of them have been independent\footnote{Naturally, with all of the authors having claimed their models to be the first ever published.}. However in the order of actual publication those have been papers of Liovic and Lakehal \cite{liovic2012}, Aniszewski et al. \cite{aniszewskiJCP} and M. Herrmann \cite{herrmann2013}.

The model of M. Herrmann is called SGSD (Sub-grid Surface Dynamics) and based on earlier efforts by Herrmann and Gorokhovski \cite{hg2009} which proposed a decomposition of small scale (subgrid) velocity. It is assumed that this subgrid field contains a specific, separable  part due to sub-grid surface tension and a model for it is proposed. Consequently, the temporal derivative of this velocity (so subfilter, surface-tension induced acceleration) is \textit{de facto} modelled using the Taylor analogy, i.e. the concept of damped oscillator \cite{MarekDM}. Even if such assumptions have been purely phenomenological, the early (purely academic) results presented in \cite{herrmann2013} are encouraging. The model however remains largely untested and, more importantly, it requires a rather complicated redistribution procedures using a much finer subgrid. Further, also the level-set distance function has to be reconstructed onto that sub-grid in the framework of RLSG (Refined Level Set Grid) method \cite{herrmann2008}. As such, for the moment it only seems suitable with a dedicated RLSG solver.

The sub-grid curvature model has been presented in the context of LEIS (Large Eddy and Interface Simulation) by Liovic \& Lakehal \cite{liovic2012}. The method stands out as historically first, even more so since its applications were signalized before the actual publication: for example a comprehensive wave dynamics study \cite{ll2011} mentions the then-yet-undisclosed modelling approach to subgrid curvature. As the subgrid normal vector is not reconstructed in this approach, $\kappa^{SGS}$ is targeted directly. It is done by a fitting procedure that tries to assess the subgrid curvature by examining a series of curvature estimates obtained through filtering the actual (grid) curvature with a succession of varying filter widths (larger than the grid size). Heuristics is used to find possible trends in this succession, which are then extended onto the subgrid scale. Example application to bubble burst is presented in \cite{liovic2012}, accompanied by previously published \cite{ll2011} estimates of under-resolved surface tension term in surf-zone turbulence simulation. While the study \cite{liovic2012} is very convincing and the model proves effective, the heuristic approach remains its weak spot, as the authors themselves admit. Also, it is immensely complicated and includes multiple conditional procedures accompanied by tabulated parametrizations within two sub-models to make the heuristics complete. That makes it extremely demanding to re-implement, although no doubt this will change thanks to continued research  \cite{liovic_fedsm_2014}.

The ADM-$\tau$ model by Aniszewski et al. \cite{aniszewskiJCP}, to which this paper is devoted is considerably simpler than both aforementioned approaches. Indeed, a deconvolution-based procedure is used to ''reconstruct'' the subgrid quantities. The publication presents multiple variants (notably ``B'' in \cite{aniszewskiJCP} and ``A'' in this paper) that either reconstruct the velocity field -- employed then to advect the interface -- or the $\nb$ and $\kappa$ fields that gives rise to the missing subgrid surface tension force directly from its definition. Since its publication, the ADM-$\tau$ has been questioned for relatively low number of published tests \cite{xiao2014}. Other researchers cited possible influence of the CLSVOF methodology \cite{ursula_private} or the inherently isotropic character of the model \cite{vincent_private}. This issues are to be addressed in the presented paper. It is naturally not our intention here to present any assessments or judgments over the published methods named above, which itself could be a topic of an interesting comparative study. Instead, the paper is devoted to continuation of work on the ADM-$\tau$ model, taking into account the fact that sub-grid surface tension modelling is no longer \textit{terra incognita}.

\section{Governing Equations}

We simulate the two-phase, incompressible flow described by Navies-Stokes equations. They will be presented here right away in the filtered form specific to the Large Eddy Simulation, since the focus of the paper is the closure for one of these subgrid terms. For the more introductory material, please refer to our previous paper \cite{aniszewskiJCP}; broader scope of information on two-phase flow equations and modeling can be found in \cite{tsz}. That said, the momentum equations have the following form:

\begin{equation}\label{eq1}
  \frac{\partial \overline{\ub}}{\partial t} + \nabla \cdot \overline{\ub}\otimes\overline{\ub}=\frac{1}{\overline{\rho}}\left(\nabla\cdot\left(\overline{\mu}\overline{\db}-\overline{p}\ib \right) + \overline{\sigma}\mbox{ }\overline{\nb}\nabla_s\cdot\overline{\nb}\delta_S \right) +\mathbf{f}_g,
\end{equation}

which is accompanied by the continuity equation:

\begin{equation}\label{eq1-2}
  \nabla\cdot\overline{\ub}=0
\end{equation}

and the advection equation for the phase indicator function:

\begin{equation}\label{eq2}
  \frac{\partial\overline{\phi}}{\partial t}+\overline{\ub}\cdot\nabla\overline{\phi}=0.
\end{equation}

In the above, $\ub$ designates the velocity vector field, $p$ stands for the pressure, $\db$ for the rate of strain tensor \[\db=\nabla\ub+\nabla^T\ub,\]  $\sigma$ is the surface tension coefficient, $\delta_S$ is the Dirac delta distribution centered on the interface $S.$ The $\nabla_S$ stands for surface differentiation operator \cite{aniszewskiJCP}. Additionally, $\ib$ is the identity tensor and $\mathbf{f}_g$ stands for body forces.

The $\phi$ will in most cases mean Level Set \cite{osher2000} distance function, itself a passively advected scalar, hence (\ref{eq2}) holds. Note that $\phi$ is used to  calculate the phase-dependent scalar fields $\mu$ and $\rho$ in (\ref{eq1}) as well as normal vectors $\nb$ and curvature $\kappa.$ The distance function is used to implicitly represent the interface (as $\phi=0$ isosurface). Additionally $\phi$  is corrected in actual implementation using the Volume of Fluid technique -- in the framework of Coupled Level-set Volume of Fluid methodology \cite{aniszewski2014caf} which improves the tracked mass/volume conservation. This however does not impair the validity of (\ref{eq1})-(\ref{eq2}).

The $\overline{(\cdot)}$ operation stands for the linear spatial filtering adopted in Large Eddy Simulation \cite{pope}. It is defined as a convolution with a compact-support filtering kernel $G$ with its specific width $\Delta$:

\begin{equation}
  \overline{u(x)}=G\star u=\frac{1}{\Delta}\int\limits^{x+\Delta/2}_{x-\Delta/2}G(\frac{x-x'}{\Delta},x) u(x')dx',
  \label{eq3}
\end{equation}

which uses the example of velocity component $u$  with the $\star$ symbol representing convolution. The $G$ operator can be applied multiple times with the same or varying widths $\Delta.$ It is a one-dimensional operator, so multidimensional filtering consists in superposition of single-dimensional filters along axes. The operation has been shown to be commutative with derivation (\cite{pope}, see eq. (\ref{commutation})) on uniform grids. Notably, if $\Delta=\Delta x,$ so that the filter width is equal to grid spacing, we may speak of trivial filtering, synonymous with ``grid-filter'' approach \cite{geurts}. In this approach, the filtering is never explicitly performed, instead the variables discretized on the grid are considered filtered with a $\Delta x$ filter width. This is the approach adopted in \cite{aniszewskiJCP} and this paper.

With (\ref{eq3}) not being distributive versus multiplication e.g. in convective term of (\ref{eq1}), filtering nonlinear expressions needs closing. (As for any quantities $a$ and $b$ describing the flow; if only $\overline{a}$ and $\overline{b}$  are known, $\overline{ab}$ is not). Thus, we can follow Labourasse et al. \cite{labourasse} and introduce a notation for closures of such term: $\tau_{ab}=\overline{ab}-\overline{a}\overline{b}.$ Additional subscripts ($l$/$r$) may appear to designate e.g. to which side of Navier-Stokes equation the closure traditionally adheres. For example,

\begin{equation}\label{eq4}
  \tau_{luu}=\overline{\ub\otimes\ub}-\overline{\ub}\otimes\overline{\ub}
\end{equation}

means that we're addressing the non-linearity of $u$ on the left-hand side of (\ref{eq1}). This term is known in LES as sub-grid stress tensor, and has been a subject of extensive research \cite{geurts,germano,pope,smagorinsky}. A term stemming from the viscous term filtration and -- to our knowledge -- not yet modeled \cite{herrmann2013} has the following form:

\begin{equation}\label{eq5}
  \tau_{l\mu D}=\overline{\mu \db}-\overline{\mu}\overline{\db}.
\end{equation}

Another term may appear if a conservative Navier-Stokes formulation is used (namely $\tau_{l\rho u}$ closing the filtration of $\rho\ub$ product in the temporal derivative).

Note that regardless of the method chosen for interface tracking, the advection equation (\ref{eq2}) for the phase indicator function $\phi$  must be closed as well:

\begin{equation}\label{eq9}
  \tau_{u\phi}=\overline{\ub\cdot\nabla\phi}-\overline{\ub}\cdot\nabla\overline{\phi},
\end{equation}

where $\phi$ should be replaced with other function e.g. when using Volume of Fluid method. The fraction function $C$ used in VOF must observe an equation analogous to (\ref{eq2}), and so a similar term (namely $\tau_{uC}$) should be considered then in place of (\ref{eq9}). Moreover, the CLSVOF method which uses both $C$ and $\phi$ simultaneously, needs \textit{both} closures. It is interesting to note that (\ref{eq9}) applies to most two-phase flow simulations (through (\ref{eq2})) as long as filtered/non-resolved velocities occur. However, no direct closures for (\ref{eq9}) have been published to this authors' knowledge.  Certain publications however have touched upon this topic in a more or less direct manner. For example, closure modelling for this ``subgrid mass transfer'' \cite{aniszewskiJCP} bears certain resemblance to Herrmann's SGSD model \cite{herrmann2013}, as the former is partly devoted to (\ref{eq2}), albeit in more conceptual way, i.e. by considering advection by subgrid velocity $\ub_{SGS}=\ub-\udash.$  Subgrid interface advection is the subject of an interesting conference paper by Liovic \cite{liovic_fedsm_2014}. Also the topic is omnipresent whenever Eulerian/Lagrangian transition is used to model sub-grid particles \cite{gaurav}.

As far as the right-hand side of (\ref{eq1}) is considered, the terms associated with the presence of the interface require closure. As explained in \cite{aniszewskiJCP} we assume $\nabla\cdot\nb=\nabla_s\cdot\nb$ as long as the $\nb$ field extends off the interface, thus considering $\nabla_s$ commutative with (\ref{eq3}) operation.
The first interface-specific term,

\begin{equation}\label{eq7}
  \tau_{r\sigma}=\overline{\nabla_s\sigma}-\nabla_s\overline{\sigma}
\end{equation}

is only non-zero if that assumption is not taken, also it requires variable surface tension coefficient (Marangoni force), which is not the case in our work\footnote{So whenever it is defined, $\sigma=\overline{\sigma}.$ For similar reasons  we do not explicitly name the terms appearing when using Favre (density-based) averaging, see \cite{vincent,labourasse} for full coverage.}. 
The second  one is $\trnn$ which stems from the product of normals and the curvature, and is defined as follows:

\begin{equation}\label{eq8}
  \trnn=\overline{\sigma}\left(\overline{\nb\nabla_s\cdot\nb\delta_s}-\overline{\nb}\nabla_s\cdot\overline{\nb}  \right).
\end{equation}

It is the very ``sub-grid surface tension'' term whose modelling is the subject of this paper.  The specific algorithm to calculate the $\trnn$ tensor will be reformulated in the following section.

Thus, the complete Navier-Stokes equations with the tensors named above, in non-conservative formulation,  would acquire the following form:

\begin{equation}\label{eq10}
  \frac{\partial \overline{\ub}}{\partial t} + \nabla \cdot \left(\overline{\ub}\otimes\overline{\ub}+\tau_{luu}\right)=\frac{1}{\overline{\rho}}\left(\nabla\cdot\left(\overline{\mu}\overline{\db}+\tau_{l\mu D}-\overline{p}\ib \right) + \overline{\sigma}\mbox{ }\overline{\nb}\nabla_s\cdot\overline{\nb}\delta_S +\trnn\right) +\mathbf{f}_g,
\end{equation}

with the closed advection equation for interface tracking:

\begin{equation}\label{eq10b}
  \frac{\partial\overline{\phi}}{\partial t}+\overline{\ub}\cdot\nabla\overline{\phi}=\tau_{u\phi}.
\end{equation}

In our work, we will assume all of the $\tau_{(\cdot)}$ tensors in (\ref{eq10}) except $\trnn$ to be zero, as the focus is on the latter.

Above discussion has not touched upon the topic of nonlinearity of the terms involving filtered density $\overline{\rho}.$  This is perhaps best visible after   (\ref{eq10}) is multiplied by $\overline{\rho}$ (the ''conservative'' formulation). An example term appearing in such scenario might be: 

\begin{equation}\label{tlrho}
  \tau_{l\rho uu}=\overline{\rho \ub \otimes \ub}-\overline{\rho}\mbox{ }\overline{\ub}\otimes\overline{\ub},
  \end{equation}
on the left hand side\footnote{The derivative $\frac{\partial\rho\ub}{\partial t}$  would also require closing.}. In a conservative formulation, this  should  be  accompanied by the term:
\begin{equation}\label{rhomass}
  \tau_{l\rho u} = \overline{\rho \ub} - \overline{\rho}\mbox{ }\overline{\ub}
  \end{equation}
in (\ref{eq1-2}). While in our work (\ref{tlrho}) and (\ref{rhomass})  are both neglected, it is interesting to note that for an interface-tracking code such as Archer3D, density is derived from distance function, so $\rho=\rho(\phi)$ (and $\overline{\rho}=\rho(\overline{\phi})).$ For that reason, we are  justified to expect that any future modelling attempts would need to consider a possible coupling e.g. between (\ref{tlrho}) and $\tau_{u\rho}$ term (\ref{eq9}) responsible for 'subgrid interface dynamics'.

%NEW=================================================================
\section{Calculation of $\tau_{rnn}$ Tensor}\label{tau_section}

\subsection{Approximate Deconvolution}

The Approximate Deconvolution technique is due to Stolz and Adams \cite{adams99,adams} and has been first  applied in context of incompressible channel flow. In this technique, the filtering equation (\ref{eq3}) is ``reversed'' in that an approximate deconvolution operator $G^{-1}$ is constructed by a truncated Taylor expansion of $G$ in $\mathbb{L}^2$ functional space. The expression is valid provided that $||I-G||<1,$ (with $I$ standing for identity operator) and has the form

\begin{equation}
G^{-1}_N=\sum\limits_{l=1}^N (I-G)^l,
\label{adm1general}
\end{equation}

where, due to truncation, $N$ is taken \cite{adams} at  $4$ or $5.$ Thus, the truncated expression for $G^{-1}$ used both in \cite{aniszewskiJCP} and here is

\begin{equation}
G^{-1}_4=4-10G+10G^2-5G^3+G^4.
\label{adm1}
\end{equation}

\begin{figure*}[ht!]
  \centering
  \includegraphics[width=0.5\textwidth]{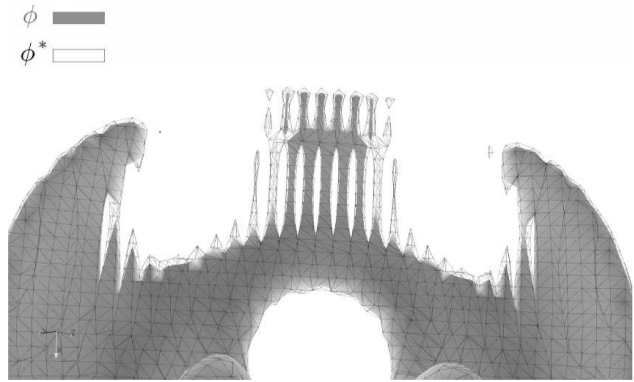}
  \caption{Deconvolution of the level-set distance $\phi$ function using the ADM \cite{adams} technique. Thin film is tracked in passive advection \cite{aniszewski2014caf}. The $\overline{\phi}$ isosurface is visible with the $\phi^*$ represented as wireframe.}\label{phi_adm}
\end{figure*}

The deconvoluted (or ``reconstructed'') variables are designated by $(\cdot )^*$ and so for example, deconvolution of the level-set distance function would have the form:

\begin{equation}
\phi^*=G^{-1}_4\star \overline{\phi}=4\overline{\phi}-10G\star \overline{\phi}+10G^2\star \overline{\phi}-5G^3 \star \overline{\phi}+G^4\star \overline{\phi},
\label{adm3}
\end{equation}
with standard iterative notation  $G^2\star \phi =G\star(G\star \phi).$  In the above, the overline designates filtering with (\ref{eq3}). More details on the technique can be found in previous publications \cite{aniszewski, aniszewskiJCP, adams, adams99}. Stolz et al. \cite{adams} have shown that the reconstructed velocity field $\ustar$ contains more spectral information than $\udash$ (high frequencies accentuation --  on the same grid, in contrast to comparable techniques using subgrids \cite{domaradzki}). As an illustration, applying  e.g. to tracking function $\phi$ in three dimensions, both original and reconstructed distributions can be seen in Figure \ref{phi_adm}.

Inspecting Fig. \ref{phi_adm} one can observe that the deconvolution has a ``sharpening'' effect which, with respect to the distance function $\phi,$ has also the effect of restoring the tracked mass/volume that was lost due to thinness of the tracked film (consistent with the legacy of ADM which originated from an image sharpening technique). Obviously, three-dimensional deconvolution is, as the filtering itself, obtained by superposition of $x,y$ and $z$ operators.

%brace yourselves!!!
\subsubsection{Filter Choice}\label{fc_sect}
\begin{figure}[ht!]
  \centering
  \includegraphics[width=0.42\textwidth]{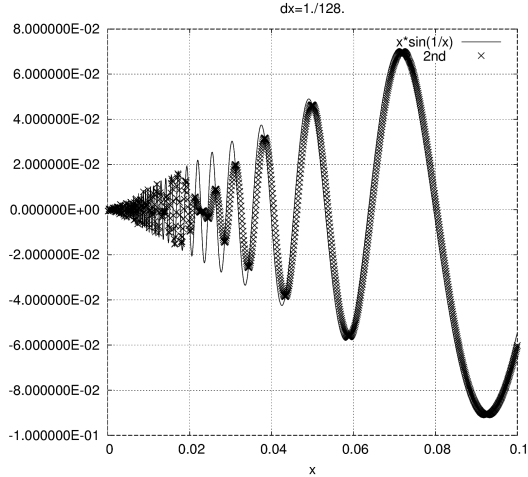}
  \caption{Example filtering using  (\ref{adm42}) of a real  $f(x)=x \sin(1/x)$ function on a $128$ point grid. The function is represented with continuous line, the filtered signal by cross symbols.}\label{jeanm}
\end{figure}

We have applied the similar $G$ filter choice strategy as in the previous publications \cite{aniszewskiJCP} and \cite{jeanmart} with small alterations adapted to differences in ADM-$\tau$ as specified later. In general, due to (\ref{eq3}) and having used the ``grid-filter'' approach, it would be desirable that the filter corresponds to the chosen spatial discretization of the deconvoluted variable. As shown by Geurts \cite{geurts} the discretization of derivatives using finite difference schemes can be shown equivalent to top-hat type filtering with the filter width equal to the discretization step. In that vein we seek the second order filter that corresponds to the second order central differencing used e.g. to calculate curvature, which as shown below, is one of the deconvoluted variables.  Therefore we have applied a second order central difference filter

\begin{eqnarray}
  \overline{f(x)} & = &f(x)+(f(x+\Delta)-2f(x)+f(x-\Delta))/4,
  \label{adm42}
\end{eqnarray}

first proposed by Jeanmart and Winckelmans \cite{jeanmart} that is characterized by its transfer function vanishing at the LES cutoff wavenumber. Its example application to a rapidly varying real function $f(x)=x \sin(1/x)$ has been illustrated in Figure \ref{jeanm}. This is the filter we have used for (\ref{adm1}) which is performed by iterating (\ref{adm42}). The actual filter choice is broader and can be made dynamically, however for the results presented below mostly (\ref{adm42}) is applied.  We will return to the ADM filter choice below when describing the new proposed ADM-$\tau$ algorithm.

\begin{figure}[ht!]
  \centering
  \includegraphics[scale=0.5]{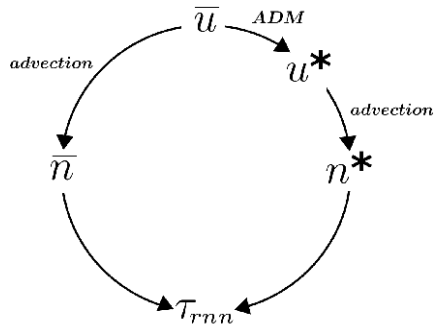}
  \caption{Diagram of the $\tau_{rnn}$ calculation algorithm, (B) approach.}\label{tau_diagramB}
\end{figure}

In the previous paper \cite{aniszewskiJCP} we have presented an implementation of ADM-$\tau$ model in variant ``B'' that is, including a single advection using the deconvoluted velocity field $\ub^*.$  We recall this in Figure \ref{tau_diagramB}. As we can see, variant B used \textit{indirect} deconvolution to obtain $\nstar$, based upon assumption that for a given velocity field $u_i(x_i,t),$ the $\mathbf{n}$ field can be represented as dependent from it. Therefore, assuming also that dependence is unique, the $\ustar$ field obtained by deconvolution was used to compute $\nstar$ which required a separate advection step. Therefore, the ``B'' variant increased the CPU cost, also in implementation, copies of $\ub,$ $\phi$ (and/or $C$) data structures must have been initialized for this purpose. 

The ADM-$\tau$ variant B is actually reminiscent of Gorokhovski-Herrmann decomposition \cite{herrmann2013} in that subgrid velocity is assumed to have acted on the interface. But in variant B, main-grid advection algorithm is directly re-applied to this subgrid velocity field, whereas the SGSD model uses this fact indirectly.

\subsection{The new calculation procedure}

However, another variant has been mentioned in \cite{aniszewskiJCP}. Variant  ``A'', is the \textit{direct} deconvolution of interface-related entities such as the normal $\nb$ and/or curvature $\kappa.$ Thus, here $\mathbf{n}^*=G^{-1}(\overline{\mathbf{n}})$ where $G^{-1}$ is the deconvolution operator . The A variant will be described now in more detailed manner, as it is the one we have decided to develop further in this paper.

\begin{figure}[ht!]
  \centering
  \includegraphics[scale=0.42]{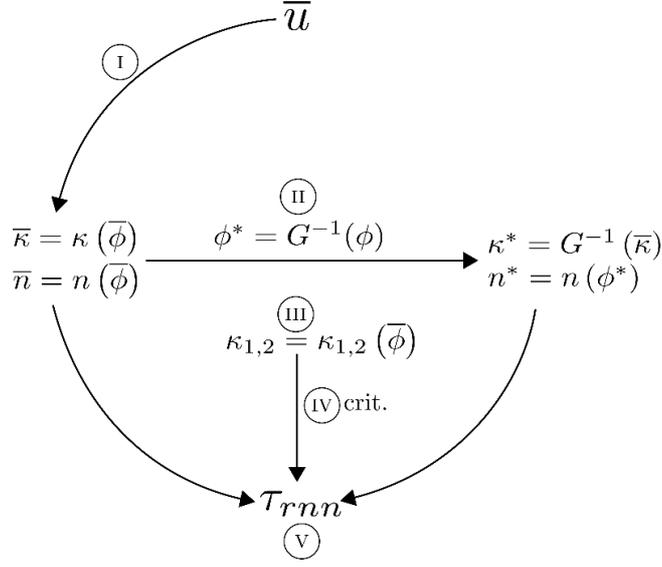}
  \caption{Detailed diagram of the $\tau_{rnn}$ calculation algorithm A variant.}\label{adm_varA}
\end{figure}

In this variant, the advection is carried out using $\udash$ (see Figure \ref{adm_varA}).  Resulting $\phi$ distance function distribution comes in turn from solution of (\ref{eq2})\footnote{As described above, CLSVOF technique is used to track the interface and as such, Volume of Fluid $C$ fraction function is also advected and used to additionally correct $\phi$ if necessary.}. At this stage, both mean curvature $\kappadash=\kappa(\phidash)$ and $\ndash=\nb(\phidash)$ are found. These two computations are not part of the model but are incorporated in most two-phase codes, as $\kappa$ is indispensable -- in one form or another -- for pressure jump computation, and $\nb$ field is required e.g. in CLSVOF context \cite{aniszewskiJCP}. Subsequently (Fig. \ref{adm_varA}, stage II) $\phidash$ is deconvoluted using an approximated $G^{-1}(\cdot)$ operator implemented via filters of choice (see \ref{fc_sect}) followed by computation of $\nstar=\nb(\phi^*).$

As for the calculation of $\kappa^*,$ we have decided to use \[ \kappastar = G^{-1}(\overline{\kappa}), \] meaning to deconvolute $\kappa$ directly, instead of implementing $\kappa^*=\kappa(\phi^*).$ This decision is mainly for implementation reasons. Naturally, for such a modification one needs to assure that \[\nabla \cdot \nstar = (\nabla\cdot\nb)^*\] both analytically and in discretized form. Indeed, it can be shown as it has frequently been done in classical single-phase Large Eddy Simulation (see \cite{pope}, p. 561), that filtering commutes with differentiation. To review this, let us write (for an uni-dimensional problem):

\begin{eqnarray}
  \label{commutation}
  \frac{\partial\ndash(x)}{\partial x} & = & \frac{\partial}{\partial x} \int G(r,x)n(x-r,t)dr  = \\ \nonumber
  & = & \int \left( \frac{\partial G}{\partial x} n + \frac{\partial n}{\partial x} G \right)dr = \\ \nonumber
  & = & \int n\frac{\partial G}{\partial x} dr + \int G(r,x)\frac{\partial n(x-r,t)}{\partial x} dr = \\ \nonumber
  & = & \overline{\left(\frac{\partial n(x)}{\partial x}\right)} + \int n(x,t)\frac{\partial G(r,x)}{\partial x} dr, \\  \nonumber
\end{eqnarray}

so that indeed,
\[\frac{\partial \ndash (x)}{\partial \xb} = \overline{\left(\frac{\partial \nb}{\partial \xb}\right)}\] provided that $\frac{\partial G}{\partial x}=0$ which is why commutation holds on uniform grids. As a double-check on an implementation level, we have further verified that for all filters tested here, the $\left( \nabla_h \cdot \nb \right)^* =  \nabla_h \cdot \nb^*$ (with $\nabla_h$ standing for discretization of differential operator) is true to machine precision even for coarse grids.

We have investigated a possibility of using a different $G$ filtering kernel for the deconvolution of the distance function, as its spatial discretization is performed using $5-$th order WENO scheme \cite{menard}. Trial runs have used a fourth order ($5-$point stencil) compact filter for $G^{-1}(\phi)$ combined with (\ref{adm42}) for $G^{-1}(\kappa).$ We have found the results to be comparable to using (\ref{adm42}) everywhere and thus it is the combination used in most simulations. Different combinations of filters, as well as their dynamic selection should be investigated in future publications.

\begin{figure}[ht!]
  \centering
  \includegraphics[scale=0.75]{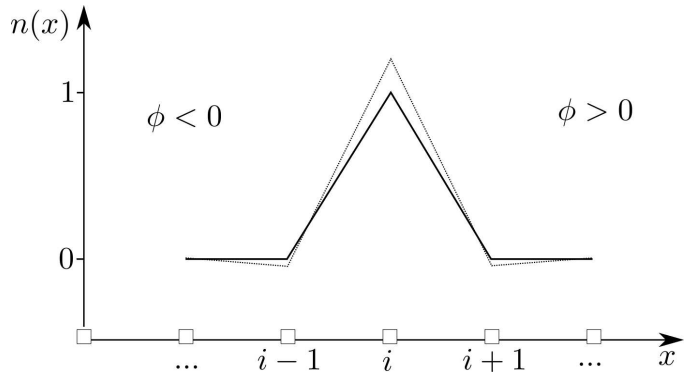}
  \caption{The deconvolution of $\nb$ normal vector field in single dimension.}\label{n_deconv}
\end{figure}

On a related note, we could consider using $\nstar=G^{-1}(\overline{\nb})$ instead of $\nstar=\nb(\phistar)$ and, in light of above arguments, similar results could be expected. Alas, $\kappa$ and $\phi$ are defined also off the interface whereas $\overline{\nb}$ is not, at least not in our CLSVOF implementation which restricts it to single cell\footnote{So that the interface always has only single cell ``thickness''. This is specific to our CLSVOF implementation whereas e.g. the THINC-type VOF  methods us more diffused interface definition \cite{xiao}.}. Taking that into account we note that applying $G^{-1}(\overline{\nb})$ along a line that cuts the interface and is perpendicular to it, would result in an accentuation of a single triangular peak. This is illustrated for $\mathbb{R}$ in Figure \ref{n_deconv}. The dotted line depicts a deconvoluted signal; we  observe the nonzero values for normal component in both nodes $i+1$ and $i-1$ which signifies a ``diffused'' interface. Summarizing, the $\left( \nb(\overline{\phi})\right)^*_h \ne \nb (\phi^*)_h$ and we settle on deconvolution of $\phi$ instead.  Note that the ``B'' variant of ADM-$\tau$ model is more suited for setups involving VOF-type interface tracking with sharp $C$ fraction function distributions, without directly deconvoluting them.

With all the necessary variables known, $\tau_{rnn}$ is computed directly from (\ref{eq8}) (Fig. \ref{adm_varA}, stage V). The restriction of $\tau_{rnn}$ to the interface is ensured by an implementation of $\delta_s$ Dirac distribution which in this case uses the distance function $\phi$:

\begin{equation}
  \label{dirac}\delta_s(\phi)=
  \begin{cases}
    \frac{1}{2}\left(1+\cos\left(\frac{\pi\phi}{f(\Delta x)}\right)\right) & \Leftrightarrow  |\phi|<f(\Delta x)\\
    0 & \Leftrightarrow  |\phi|>f(\Delta x),
  \end{cases}
\end{equation}

where $f(\Delta x)$ is a sharpness parameter usually set to $\Delta x/4.$

ADM in classical, single phase implementations \cite{adams99} is accompanied by an additional relaxation term (added to right-hand-side of (\ref{eq1})) ensuring proper energy transfer to the undresolved scales. This additional dissipation can also by obtained by filtering $\udash$ with a secondary filter with kernel equal to $G^{-1}\star G$. This technique can be used with ADM-$\tau$ variant B \cite[p. 7375]{aniszewskiJCP}. However, in context of the proposed A variant, in which distance function $\phi$ and curvature $\kappa$ are directly deconvoluted, application of secondary filter to any of the two would need additional justification. It would most likely directly influence the interface geometry\footnote{While the influence of ADM-$\tau$ is indirect,  via $\trnn$ force.}  (by altering $\phi$) and would in effect constitute a closure to (\ref{eq9}) coupled to (\ref{eq8}). This seemingly interesting perspective -- which could be described e.g. on the ground of multiscale analysis \cite{cd_part1} that utilizes the concept of ``scale transfer'' in liquid system, akin to kinetic energy transfer in single-phase turbulence -- will be a subject of further research. For the moment, it is possible that ADM-$\tau$ variant A recovers only a part of the nonresolved scales' influence on subgrid surface tension, and could be assisted by accompanied by additional models.

\subsection{DNS convergence criteria}

As

\begin{equation}\label{cvrg}
  \lim\limits_{\Delta x \to 0} \nstar=\nb,\mbox{ and } \lim\limits_{\Delta x \to 0} \phistar=\phi,
\end{equation}

the model converges to DNS and $|\tau|\to 0.$ However, we are interested in having more control over this convergence process, thereby making sure, that regardless of e.g. the choice of the $G^{-1}$ operator used for deconvolution, or the model variant choice (A or B), zero values of $\tau_{rnn}$ will be produced provided that all variables in (\ref{eq8}) are resolved. Other motivation is that this additional degree of control may let us adapt the values of resultant force to the topology of interfacial surface in a more diversified manner, as will be shown further.

Thus, a new ADM$-\tau$ feature this paper introduces is the forced 'DNS convergence' criteria. They rest upon the assumption that the model should yield zero $\tau_{rnn}$ values for well resolved flow. Since the model computes in fact a force resulting from the emergence of non-resolved (sub-grid) curvatures, we introduce a criterion which, conceptually, restricts it to act only where local curvature cannot be resolved. As both normal vectors $\nb$ and curvature $\kappa$  are found estimating first and/or second derivatives of  a chosen interface-description function (e.g. $\nabla \phi$ or $\nabla C$), the fixed grid size $\Delta x$ is limiting interface curvature radius to which we could consider the $\nb$ 'resolved'.  Thus, for curvatures

\begin{equation}\label{kappamax_def}
  |\kappa| > \frac{1}{ \Delta x}
\end{equation}

normal vector approximation is feasible\footnote{Which does not mean that basic centered finite differencing will yield proper $\nb$ components in all  cases \cite{aniszewski}}. Therefore, we expect

\begin{equation}\label{tauzero}
  |\trnn | > 0 \Leftrightarrow |\kappa| \ge \frac{c_{\kappa}}{ \Delta x} \colon = \kappa_{max},
\end{equation}

while the exact limiting value would depend on model constant $c_{\kappa}.$  By default $c_\kappa=1$ which is equivalent to (\ref{kappamax_def}) however smaller values will loosen the restriction, down to $c_\kappa=0$ which would effectively turn the criterion off. The additional degree of freedom this introduced may prove very useful e.g. to selectively scale the resultant subgrid surface tension force depending on flow region/character.

%smoothing description, put Heaviside

To ensure smoothness of solutions including the modeled $\trnn$ tensor, we apply a smooth jump function to (\ref{tauzero}). It is a regularized version of the Heaviside function \cite{towers2009}:

\begin{equation}\label{heavymetal}
  H^{\epsilon}(x)=
  \begin{cases}
    H(x) & \Leftrightarrow  |x|\ge\epsilon \\
    \frac{1}{2}+\frac{x}{2\epsilon}+\frac{sin(\pi x/\epsilon)}{2\pi}     & \Leftrightarrow  |x|\le\epsilon,
  \end{cases}
\end{equation}
where $\epsilon=O(\Delta x)$ and $H(x)$ is a widely used, discontinuous Heaviside function $H(x)=\frac{1}{2}\left( sgn (x) +1\right)$. In actual implementation we want to obtain a rather broad smoothing margin in curvature space and so $\epsilon$ is increased as to be of order of $\kappa_{max}$. For example,  results presented in Section \ref{osc_sect} use $\epsilon=\frac{\kappa_{max}}{5}.$ Having defined the function (\ref{heavymetal}), we put:

\begin{equation}\label{crit5}
  \trnn=\trnn H^{\epsilon}(c'_{\kappa}\kappa-\kappa_{max}),
\end{equation}

with $c'{\kappa}$ being another scale coefficient possibly equal to unity.

\begin{figure}[ht!]
  \centering
  \includegraphics[scale=0.42]{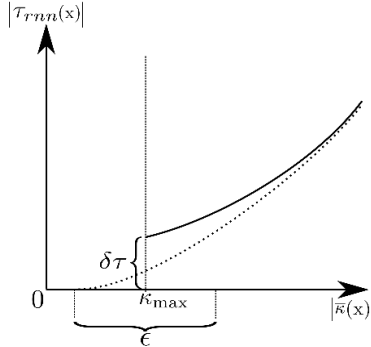}
  \caption{The smooth (dotted line) and abrupt (continuous) transitions of $\trnn$ magnitude in curvature space.}\label{crit2}
\end{figure}

Including the (\ref{crit5}) restriction imposes that in  presence of well-resolved curvatures $|\trnn|=0.$ Dropping the smoothing function of (\ref{crit5}) is also possible. By analogy to spectral analysis, the Heaviside expression in (\ref{crit5}) may be seen as a transition function in curvature space (see Figure \ref{crit2}). Using larger values of $\kappa_{max}$ may -- considering the fact that $\nb^*\rightarrow \nb$ as the grid is refined -- lead to both curves in Figure \ref{crit2} be very close to each other, and/or the jump value $\delta \tau$ be very small.

\subsection{Criteria and Anisotropy}\label{crit_sect}
%Enhancement to principials (separately)
The inherent characteristic of the ADM-$\tau$ model is that it depends on the mean curvature of the interface, so that manifolds characterized by similar \textit{mean} curvatures will be treated just the same. To address this, in imposing above criteria, principal local surface curvatures $\kappa_1$ and $\kappa_2$ can be considered. This has been implemented as a next stage of ADM-$\tau$ model development, taking the form of a following constraint:

\begin{equation}\label{crit1}
  \trnn >0 \Leftrightarrow \left(\kappa_1 > \kappa^1_{max} \wedge \kappa_2 > \kappa^2_{max}\right),
\end{equation}

with $\kappa^i_{max}$ being the (possibly identical) limiting $\kappa_i$ values similar to  $\kappa_{max}.$

Obviously,  such a formulation requires computing both  principal curvatures beforehand. However, the benefit lies in non-isotropic model behaviour thus obtained. For example, elongated liquid ligaments , whose longitudinal curvature is small, while transverse is significant, are now not affected by the model, while spheroidal blob of comparable radius are. This addresses one of major shortcomings of the original model formulation \cite{aniszewskiJCP}.

While the level-set function $\phi$ allows to extend the $\nb$ field off the interface, the principal curvatures are found only in its strict vicinity to limit the computational cost. They are calculated as eigenvalues of the local geometry tensor

\begin{equation}\label{crit3}
  G=\nabla\nb^T\left(\mathbf{I}-\nb\nb^T\right),
\end{equation}

preceded by computing the Hessian $H(\phi)$ matrix which partly constitutes the first term of (\ref{crit3}):

\begin{equation}\label{crit4}
  \nabla \nb^T = \frac{\left(\mathbf{I}-\nb\nb^T\right) H(\phi)}{|\nabla\phi|}.
\end{equation}

The eigenvalues are extracted using matrix invariants. This procedure has previously been applied e.g. to find curvatures of implicit surfaces in context of image processing \cite{preusser} and can readily be applied to $\nb=\nb(\phi)$ \cite{deschamps}. Note that $\trnn$ is only defined in the global Euclidean $(X,Y,Z)$ basis and we only use \textit{values} of $\kappa_i$ to help trigger the model. We do not, however, use the eigen\textit{vectors} of $G$ that constitute the principal directions of curvature \cite{preusser}. Expressing the $\trnn$ in the basis of curvature directions and normal seems a very interesting perspective and should be an object of further research.

\begin{figure}[ht!]
  \centering
  \includegraphics[width=\textwidth]{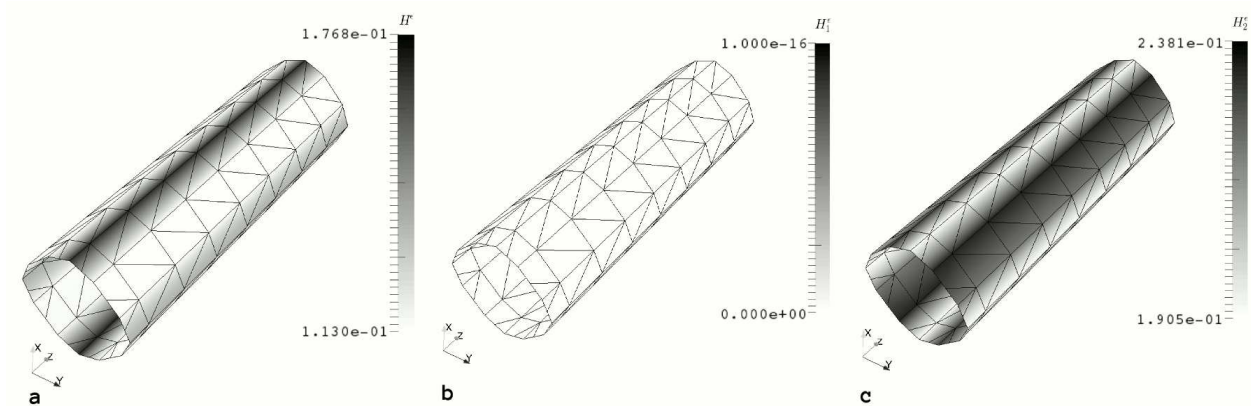}
  \caption{Values of smoothed Heaviside functions (\ref{heavymetal}) plotted for a cylinder case on coarse grid. (\textbf{a}) Isotropic criterion, $H^\epsilon$ function. (\textbf{b,c}) Non-isotropic criterion: functions $H^\epsilon_{1,2}$ calculated using $\kappa_{1,2}$ respectively.}\label{crytlinder}
\end{figure}

By an analogy to (\ref{kappamax_def}) and (\ref{tauzero}), criterion (\ref{crit1}) is implemented using smoothing functions $H^\epsilon_{1,2}$ that utilize principal curvatures $\kappa_{1,2}$ respectively and have the form (\ref{crit5}). We present a computational example using a cylinder-shaped surface. This particular cylinder has been initialized in such a way, that the $d\approx 4\Delta x,$ where $d$ is the cylinder's diameter and $\Delta x$ is the uniform grid mesh size. As such, the mean curvature $\kappa$ is large enough to approach the limiting value of $\kappa_{max},$ so that $H^\epsilon(\kappa-\kappa_{max})>0;$ its actual values  are visible in Figure \ref{crytlinder}a. At the same time, only the second of principal curvatures is non-zero, so that the distribution of $H^\epsilon_2$ in Fig. \ref{crytlinder}c is clearly different from that of $H^\epsilon_1$ pictured in Fig. \ref{crytlinder}b\footnote{Note that legend values for Fig. \ref{crytlinder}b are of order of $10^{-17}.$}. Thus, by replacing (\ref{crit1}) with

\begin{equation}\label{crit12}
  \trnn >0 \Leftrightarrow H^\epsilon_1>0 \wedge H^{\epsilon}_2>0,
\end{equation}

we cause $\trnn$ to vanish in situations such as Figure \ref{crytlinder}b. The actual implementation of the condition leaves some freedom to the programmer. Above condition may be used (accompanied by $H$ functions used to smooth $\trnn$),  but it is just as well possible to put:

\begin{equation}\label{crit52}
  \trnn=\trnn H^{\epsilon}_1(c^1_{\kappa}\kappa_1-\kappa_{max})H^{\epsilon}_2(c^2_{\kappa}\kappa_2-\kappa_{max}),
\end{equation}

which will imply condition (\ref{crit12}) rescaling the resulting force slightly. The role of the coefficients $c^{1,2}_\kappa,$ in (\ref{crit52}) which are by default set to $1,$ will be  explained further.

It is interesting to observe the influence of which approach (isotropic/non-isotropic) is chosen in the above. We have included such discussion in context of actual simulations in Section \ref{sephase_sect}.

As was stated in the beginning of this section, the purpose of the introduction of the criteria  (\ref{crit12}) is not the DNS convergence itself. The latter stems from (\ref{cvrg}) and has already been demonstrated for previous ADM-$\tau$ model versions (which used no criteria, e.g. Fig 23 in  \cite{aniszewskiJCP}). It can also be inferred from observing filter (\ref{adm42}) behaviour as presented in Fig. \ref{jeanm}: we see that for $x>\Delta x\approx 0.0357$ the filtering is trivial and $f(x)=\overline{f(x)},$ suggesting devonvolution converges once this filter is used.

Instead, the purpose of the presented criteria mechanism is to make the model anisotropic. In complex flows, it is easy to find situations in which one of the principal curvatures $\kappa_{1,2}$ is resolved while other is not. This occurs not only on surface of ligaments but also at the edges of liquid sheets or surface of initially flat film undergoing wrinkling. Thus, in such situations it is possible that e.g. $\overline{\kappa_1}=\kappa_1$ while the same does not hold for $\kappa_2.$ It would be then profittable to deconvolute the principal curvatures separately; however, at this stage of ADM-$\tau$ development we have not included such possiblity. Instead, we aim  at deconvolution of mean curvature, computing the $\trnn$ according to algorithm presented in Fig. \ref{adm_varA}. We later redistribute the resultant force according to the magnitude of resolved principal curvatures. Though not equivalent to treating principal curvatures completely separately, the technique proves very efficient as will be shown below.

In (\ref{crit52}) $\epsilon$ stands for the smoothing margin in curvature space, as discussed in the context of (\ref{heavymetal}). Its role is to smoothen (or scale) the resulting $\trnn$ distribution. Small values result in less $\trnn$ prevalence, as it will be abruptly zeroed once local curvature falls below threshold value. The action of $\epsilon$ applies to both principial curvatures.

Coefficients $c^1_\kappa$ and $c^2_\kappa$ are to be understood in curvature space as well. Set to $1,$ they  result in both principal curvatures being treated (weighted) identically in (\ref{crit52}). This is the approach chosen for the work presented in this paper. However, $c^i_\kappa$ may provide additional degree of freedom in the design of the criterion.  In order to employ these coefficients in non-trivial manner, we need a way to discern between principal curvatures (for example by sorting them and assigning indices in such a way, that $\kappa_1<\kappa_2$).  Then, values of products $c^1_\kappa\kappa_1$ and $c^2_\kappa\kappa_2$ could -- by balancing the values of coefficients --  be made arbitrarily close. In such way, the criterion can be made more isotropic (if $|c^1_\kappa\kappa_1-c^2_\kappa\kappa_2|$ decreases) or anisotropic (if that difference is increased). In other words, $c^1_\kappa$ and $c^2_\kappa$ in (\ref{crit52}) consitute model isotropy coefficients. This can be used to locally modify model isotropy in a dynamic manner and will be subject of further research. 

%=======================================================================================
\section{Numerical Simulations}

\subsection{Description of the Archer 3D code}

The \textit{Archer3D} code is a Marker-and-Cell (MAC) type (staggered grid) solver for Navier-Stokes equations, originally developed by S. Tanguy \cite{Tanguy2005} and continued by M\'{e}nard et al. \cite{menard}.  The momentum equations are solved using the projection method, with Poisson equation solution augmented by preconditioned MGCG (Multigrid-Conjugate Gradients \cite{brandt}) technique. Temporal discretization is performed by means of $3$rd order Runge-Kutta scheme. Spatial discretization uses WENO \cite{shu} with several variants available as switchable options.  One-fluid approach (single Navier-Stokes equations set with jump at the interface \cite{tsz}) is used for two-phase flow modelling. The Level-Set (LS) method \cite{osher88} is used to implicitly track the interface, and simultaneously VOF \cite{hn} method helps both to conserve traced phase mass/volume, and provides the explicit interface location. These interface tracking techniques coexist in the framework of CLSVOF \cite{sussman1,aniszewskiJCP} method. The Ghost Fluid (GFM) method \cite{fedkiw-gfm} is employed to enable numerically stable pressure jump calculation. The solver is written in FORTRAN 90 and fully parallelized using MPI subroutines, tested on up to 8192-core configurations \cite{vaudor}. 

Archer3D  provides for ``classical'' LES single-phase closure models \cite{chesnel2011-1} for $\tau_{luu}.$ However as said above, for the presented simulations, to avoid any mutual influences between these standard subgrid stress models and the tested $\trnn$ model, the former had all been disabled and  so $\tau_{luu}=0$ is assumed. This allows for more reliable comparison with DNS results, as only the influence of ADM-$\tau$ is taken into account.

\subsection{Oscillating droplet}\label{osc_sect}
\subsubsection*{``Super-coarse'' grid}
The first presented test case aims at testing both the applicability of the ADM-$\tau$ model, effectiveness of the criterion (\ref{crit1})  and preliminarily assesses the influence of sub-grid surface tension on a physical phenomenon. We simulate, in three dimensions, a droplet of water ($\rho_l=1000$kg/m$^3$, $\mu_l=1.79\cdot 10^{-3}$ Pa$\cdot$s, $\sigma=0.074$ N/m) suspended in second phase that has the characteristics of gas, so $\mu_g=1.7\cdot 10^{-5}$ with density raised to $10$kg/m$^3$. The density change does not impair the results in any way (at least in context described below) and alleviates any possible momentum conservation issues. The domain is a box\footnote{The same configuration has been used in \cite{aniszewski2014caf} except for the droplet radius, as explained further.}  with $L=0.1$m, there is no gravity. We set the droplet radius to $r=1.63$cm. The idea of the test is to investigate the droplet which is barely resolved on a coarse grid (just over $4$ grid-cells in diameter) and thus, a uniform grid of $16^3$ points is used.

Lamb \cite{lamb, aniszewski2014caf} provides the description of expected frequency and amplitude decay for an oscillating three-dimensional droplet. For the frequency $f,$ we can put

\begin{equation}\label{freq}
  2\pi f=\sqrt{\frac{8\sigma}{\rho_l r^3}},
\end{equation}
where $r$ is the droplet radius (in meters), while the decay rate of the $n$-th mode is given by

\begin{equation}\label{osc_decay}
  a_n(t)=a_0e^{-t/2\pi\tau_n}, \mbox{ where } \tau_n=\frac{r^2}{(n-1)(2n+1)\nu_l}.
\end{equation}

For the given configuration the maximum computed curvature on the droplet surface is $\max_{t=0}(\kappa)=131$ m$^{-1}$, which is about $80\%$ of maximum resolvable curvature $\kappa_{max}^{16^3}$ for this grid-step $\Delta x$ size:
\begin{equation}\label{kappamax16}
  \kappa_{max}^{16^3}=\frac{1}{\Delta x}=160.
\end{equation}

Nevertheless, due to the fact that the $\tau_{rnn}$-application criterion (\ref{crit5}) is smoothed in curvature space over the width of $\frac{\kappa_{max}^{\cdot}}{5},$ we have  \begin{equation}\label{kappadif}\max_{t=0}(\kappa(r))>\frac{4}{5}\kappa_{max}^{16^3}\end{equation} thus engaging the ADM-$\tau$ model. We note that the difference between the sides of (\ref{kappadif}) is small, so only relatively small magnitudes of $\tau_{rnn}$ should be expected.

At the same time, the problem  stays within the scope of the linear instability theory, as  initial droplet deformation is $\epsilon=0.05r.$   We can thus  apply (\ref{freq}) and expect $f\approx 1.860$ oscillations per second for this $r.$

\begin{figure}[ht!]
  \centering
  \includegraphics[scale=0.84]{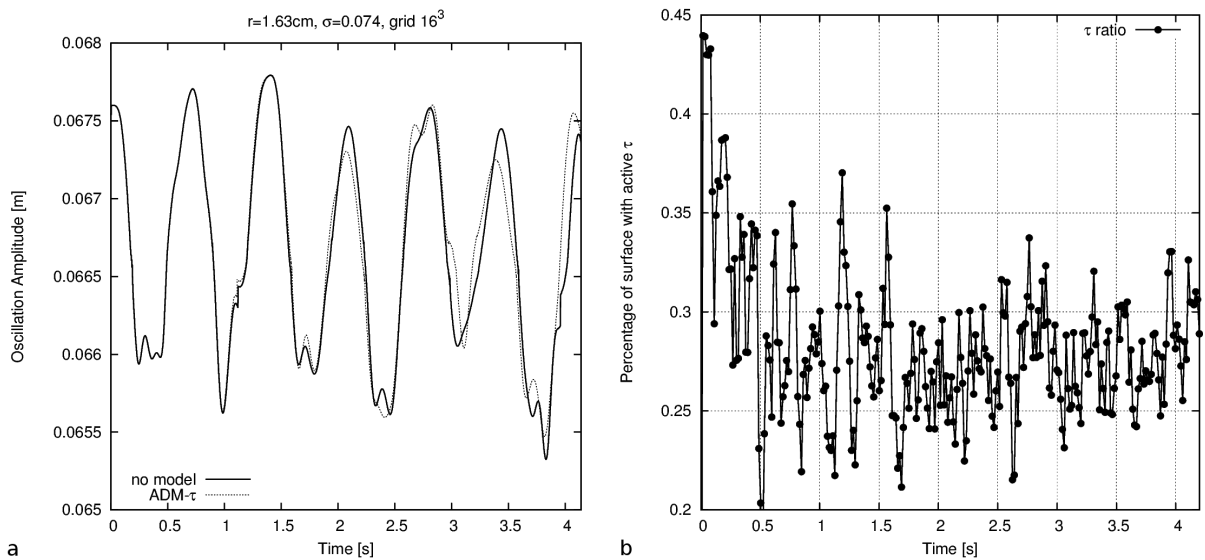}
  \caption{Droplet oscillation, (\textbf{a}) Oscillation amplitude, no model (solid line), ADM-$\tau$ (dashed line); (\textbf{b}) percentage of interface where the model was activated.}\label{osc_amp1}
\end{figure}

In Figure \ref{osc_amp1}a  temporal evolution of oscillation amplitude is presented, as obtained using a $16^3$ computational grid. Clearly, Archer severely under-predicts this frequency on $16^3$ grid, with the droplet of $r=1.63$cm being barely resolved. As a matter of fact, a droplet with $r=1.60$cm would already fall below the resolvability limit as it would have $4$ grid-cells in diameter making it impossible to properly apply a WENO/VOF ($5$-stencil) based code \cite{popinet2}. First amplitude peak in Fig. \ref{osc_amp1}a is expected at $0.53$s while it occurs at $0.72$s. \footnote{Archer results for this case on more refined grids, show that both frequency and amplitude decay predictions are correct, see \cite{aniszewski2014caf} (more specifically: Figure 10, \textbf{M4} curve). }

Using ADM-$\tau$ we see clearly in Fig. \ref{osc_amp1} that a shift is introduced in amplitude peaks, amounting to slight frequency increase. This is attributed to the increase in resolved $f_\sigma$ and thus  $\sigma$ in (\ref{freq}), and resembles an effect one would expect in a parametric study when increasing $\sigma.$ Figure \ref{osc_amp1}b portrays the percentage of the interfacial surface where non-zero values of $\tau_{rnn}$ occurred. As we can see, the median value is between $25$ and $30$ percent marks. At the same time, tensor magnitudes of the order of $10^{-4}$ (never exceeding $10^{-3}$) were recorded for this simulation.

Obviously, a more prolate initial droplet shape could have been considered as to increase the initial curvature and thus obtain higher $\tau_{rnn}$ magnitudes, however we could not be confident in (\ref{freq}) prediction at that time.

\begin{table}
  \begin{center}
    \caption{Cumulative peak shifts in first five amplitude peaks (seconds), cumulative amplitude shifts (absolute)  and respective $\trnn$ surface ratios for varying smoothing width $\epsilon$ in oscillation test case.}
    \begin{tabular}{c|c|c|c}
      $\epsilon$ as  $\frac{\kappa_{max}}{\cdot}$  &   $\Sigma df$ (s)  & $\Sigma |dA|$ ($m$)  &  $\trnn$ ratio ($\%$) \\ \hline
      $10.0$   &   $-6.660\cdot 10^{-2}$  &   $3.457\cdot 10^{-4}$ &  $7\%$   \\
      $7.6$   &   $-4.331\cdot 10^{-2}$  &   $4.271\cdot 10^{-4}$ &  $13\%$   \\
      $5.0$   &   $1.627\cdot 10^{-1}$  &   $3.794\cdot 10^{-4}$ &  $26\%$   \\
      $4.0$   &   $1.623\cdot 10^{-1}$  &   $8.676\cdot 10^{-4}$ &  $40\%$   \\
      $2.0$   &   $1.311\cdot 10^{-1}$  &   $8.605\cdot 10^{-4}$ &  $46\%$   \\
      $1.0$   &   $1.421\cdot 10^{-1}$  &   $1.102\cdot 10^{-3}$ &  $47\%$   \\
    \end{tabular}\label{osc_tab}
  \end{center}
\end{table}

%[table description longer?]
The simulation illustrated with Fig. \ref{osc_amp1} is described quantitatively in the third row of Table \ref{osc_tab}, of which each row corresponds to another simulation. All of the simulations depicted in Table \ref{osc_tab} are the same configurations, except that the smoothing margin $\epsilon$ of (\ref{heavymetal}) is increased (leftmost column, from $\kappa_{max}/10$ to $\kappa_{max}$) which results in the ADM-$\tau$ model being activated in more computational cells which is visible in the rightmost column as a percentage of total droplet surface. Second column of the table is a sum -- over first five amplitude peaks -- of time shifts caused by the model. The sum is signed: the shifts to the left (meaning frequency increase since the oscillation period shortens) are taken with positive sign. %[MORE: IT IS TIMES, NOT FREQ]
Finally, the third column has a sum of amplitude shift modules.

It is noticeable that the expected frequency shift is most visible when $\epsilon=\kappa_{max}/5$ is selected for (\ref{heavymetal}). This parameter choice makes, through application of (\ref{heavymetal}), the model active in about $26\%$ of grid-cells, also the amplitude change $\Sigma |dA|$ is second-smallest for this $\epsilon.$ This is seen as desirable, since the amplitude decay rate (\ref{osc_decay}) does not depend $\sigma.$ Interestingly, we see that setting $\epsilon=\kappa_{\max}/10$ (Table \ref{osc_tab} row 1) yields an even smaller amplitude deviation. However, such value also causes an actual decrease of oscillation frequency -- as seen in the second column -- and may be due to the model being active in only $7\%$ (averaged) of the interfacial grid-cells, which accounts for one 'pulse' at the beginning of the simulation once the curvature near the top of the droplet is maximum, with the $\trnn$ being largely zero ever after.

To reiterate  -- while the tests presented in this section may at the first glance seem of little relevance to the complex problems encountered in simulations of two-phase flow, since only singular droplet is simulated, using a computational grid with unusual coarseness -- the issue at hand is the behaviour of a droplet at the local grid's resolvability limit. Thus, similar mechanism is met locally even in complex problems such as atomisation of complex mixing, or more generally whenever substantial fragmentation occurs, as eventually non-resolved fluid parcels are bound to appear.

\subsubsection*{Coarse grid}

\begin{figure}[ht!]
  \centering
  \includegraphics[scale=0.84]{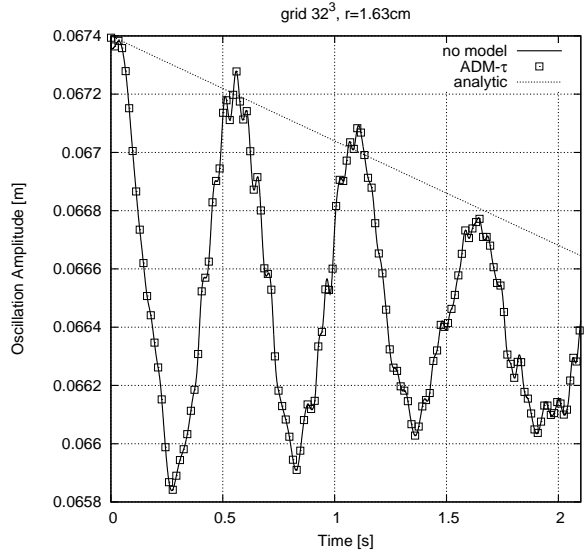}
  \caption{Droplet oscillation amplitude of a droplet with radius $r=1.63$cm using a $32^3$ grid. Series for ADM-$\tau$ (squares), simulation without a model (solid), analytic decay curve $a_2(t)$ using (\ref{osc_decay}).}\label{32osc_amp_decay}
\end{figure}

An example of the droplets reaching the non-resolvability limit on a different grid is given in this subsection. Using a slightly more refined, uniform  grid with $32^3$ cubic elements for the exact same simulation (keeping $r=1.63$cm), we arrive at the amplitude decay graph which differs from Fig. \ref{osc_amp1}a as the flow is much better resolved. Inspecting the Figure \ref{32osc_amp_decay}, we see that, to begin with, the amplitude decay actually takes place, as the peaks' decrease obeys -- albeit roughly --   the analytic decay curve (\ref{osc_decay}) plotted for the dominating second mode. This may be attributed to the viscous terms in governing Navier-Stokes equation being much more resolved (using this grid, we get $r=0.0163m\approx 5.21\Delta x$). There is no perceptible difference between the 'no model' and ADM-$\tau$ curves, as (owing to the criterion (\ref{crit5})) the model is almost striclty off for this droplet radius.

\begin{figure}[ht!]
  \centering
  \includegraphics[scale=0.84]{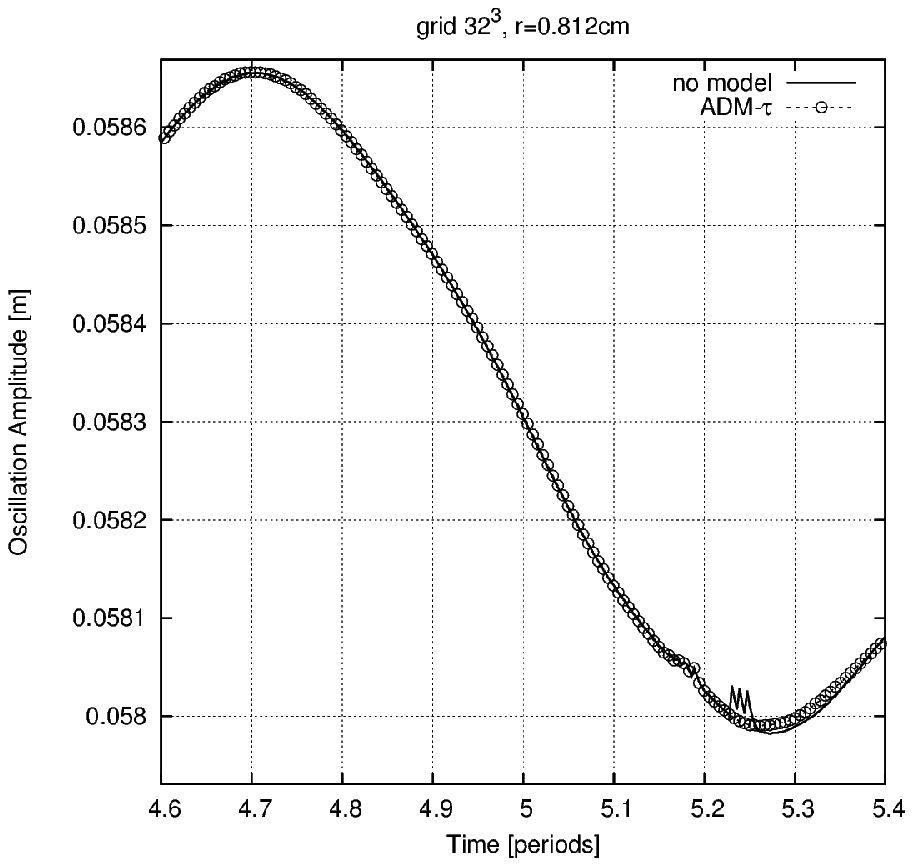}\includegraphics[scale=0.84]{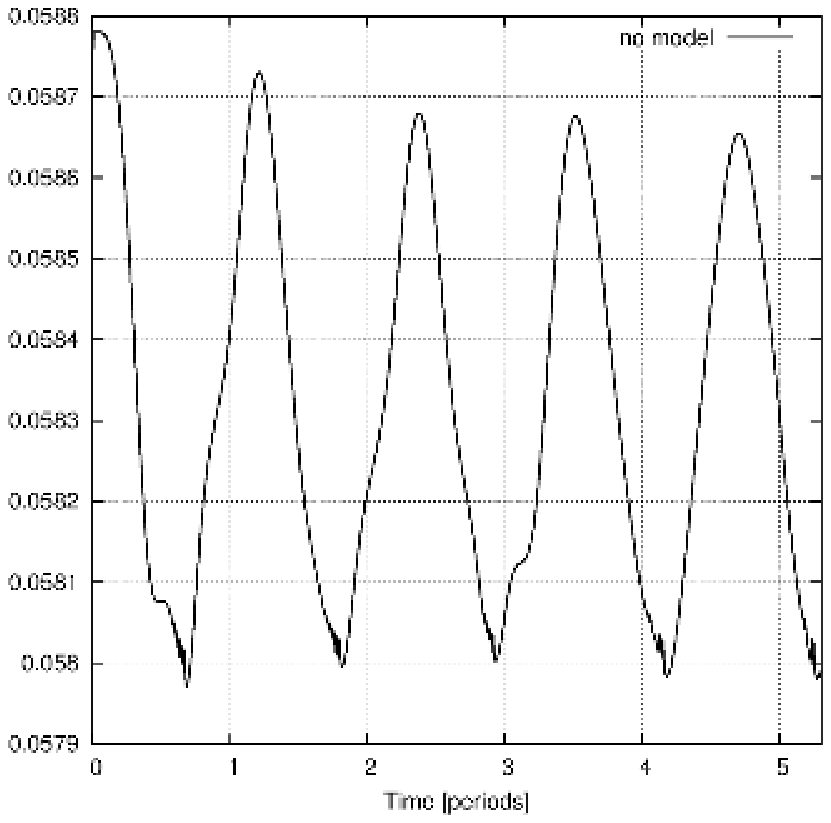}
  \caption{(\ref{osc_decay}). (\textbf{a}) Oscillation amplitude for droplet radius $r=8.12$mm. Curves for the ADM-$\tau$ (dashed circles) and no model (solid). (\textbf{b}) Amplitude decay for the same simulation, simulation without the model.}\label{32influ}
\end{figure}

Once the radius is made smaller, (\ref{crit5}) is fulfilled, e.g. for half the above radius, that is $r=0.81$cm the model contributes at around $22\%$ ratio when $\epsilon$ is set to its standard value of $\kappa_{max}/5,$ similar to the third row in Table \ref{osc_tab}. However, the $\Sigma df \approx 8\cdot 10^{-3}$ for that simulation, and the model influence is substantially smaller. This is visible in Figure \ref{32influ}a, with barely legible shift visible towards $t=5,$ albeit the model still delivers the actual increase of otherwise underpredicted frequency\footnote{The time in this figure is scaled by the theoretical period of the second node.}. This seems to be caused by the different flow physics stemming from decreased droplet radius and overall frequency increase compared to $r=1.63$cm droplet. In this flow, the amplitude peaks (Figure \ref{32influ}b) decrease more regularly than in $16^3$ simulation (Fig. \ref{osc_amp1}a) although they no longer conform to (\ref{osc_decay}).

From a numerical point of view, simulations presented in Fig. \ref{32influ}b and \ref{osc_amp1}a differ in that more gridpoints surround the interfacial cells, changing the character of kinetic energy dissipation, which could play a role in making the $\trnn$ influence on overall simulated amplitude much less pronounced. That is not to say that model inclusion exerts no influence for $r=0.81$cm case; for example if one considers the maximum calculated velocity magnitude recorded (per iteration) in the domain:

\begin{equation}\label{osc_maxvel}
  \max\limits_{i,j,k\in\lb 1,N\rb}|\ub_{ijk}(t^{n})|,
\end{equation}

we obtain up to $4\%$ of difference between two (model/no model) simulations plotted in Fig. \ref{32influ}.

\subsubsection*{The influence of CLSVOF methodology}

Even if the oscillating droplet test-case is moderately uncomplicated from a numerical point of view, it has been shown to be useful and demanding in validating e.g. curvature and normal vector calculation schemes \cite{popinet1, aniszewski2014caf}; on the other hand, number of authors focus on the accuracy of the kinetic energy dissipation \cite{vaudor}. The CLSVOF method \cite{sussman2006,aniszewski,menard} has been extensively tested in this context \cite{aniszewski2014caf}; still, as involves manipulating the values of $\phi$ in the interfacial cells -- however carefully would it be carried out -- questions arise as to whether specific disturbances could be introduced on the surface \cite{aniszewski2014caf}, disrupting the curvature calculation and thereby indireclty undermining the ADM-$\tau$ predictions. Therefore, we propose a test with the purpose of investigating such possiblity. This will be carried out by using the ADM-$\tau$ model in conjuction with the ''pure'' Level Set method, with no coupling with Volume of Fluid. This is easily implemented since the ADM variant presented here involves only the $\phi$ function, and doesn't depend on $C$ fraction function in any way.

We use a similar setting as mentioned above (including the $16^3$ grid), whereas the droplet radius has been increased to $3$cm. This is due to the fact that smaler radii cannot be reliably resolved using LS and a $16^3$ grid, as all the tracked fluid volume is quickly lost. We modify (\ref{tauzero}) by setting $c_\kappa=0$ so that the model is applied everywhere; this in turn is due to the $r=0.03$m droplet having smaller curvature than the values given in (\ref{kappamax16}) and (\ref{kappadif}) for smaller radii.

\begin{figure}[ht!]
  \centering
  \includegraphics[scale=0.84]{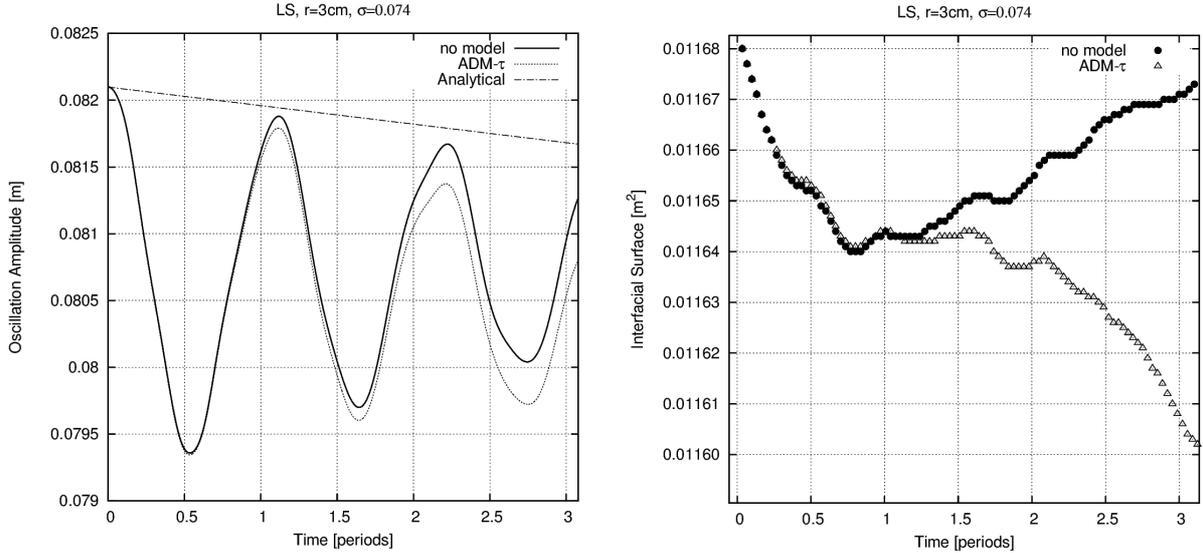}
  \caption{Simulated oscillation amplitude (left) and interfacial surface (right) using the Level-set method with and without the ADM-$\tau$ model.}\label{osc_ls}
\end{figure}

Once again we investigate the droplet oscillation, which for the $r=3$cm droplet should have a frequency of approximately $0.74$ s$^{-1}.$ The results are shown in Figure \ref{osc_ls}. Time has again been scaled by the predicted oscillation period. It is visible, that the oscillating motion has been resolved relatively well which stems from the fact of droplet being larger than the $r=1.63$cm presented before. As we can see, Archer overpredicts the oscillation period for both curves, however, as previously when CLSVOF was used, inclusion of the ADM-$\tau$ causes a shift -- this time visible with naked eye -- of the amplitude curve in the direction of frequency increase. The sum of peak shifts is $2.65\cdot 10^{-1}$s is of the comparable order that values presented in Table \ref{osc_tab} for CLSVOF method.

This desired result proves that the model's prediction is consistent with those obtained previously. The right-hand side plot of Fig. \ref{osc_ls} shows that the computed interfacial surface is slightly lessened when using ADM-$\tau,$ although the difference is of the order of $10^{-5}$ m$^2.$ This interesting variance could contribute to the perceived lower oscillation amplitudes when the model is used (dashed line), although the speciffic mechanism of that requires further investigation.

%[I THINK WE'RE NOT IN KANSAS ANYMORE!]

%\include{section43}

\subsection{Separation of Phases}\label{sephase_sect}
\subsubsection{Introduction to the test-case}
%[I can see a number of plots already prepared]

\begin{figure}[ht!]
  \centering
  \includegraphics[width=\textwidth]{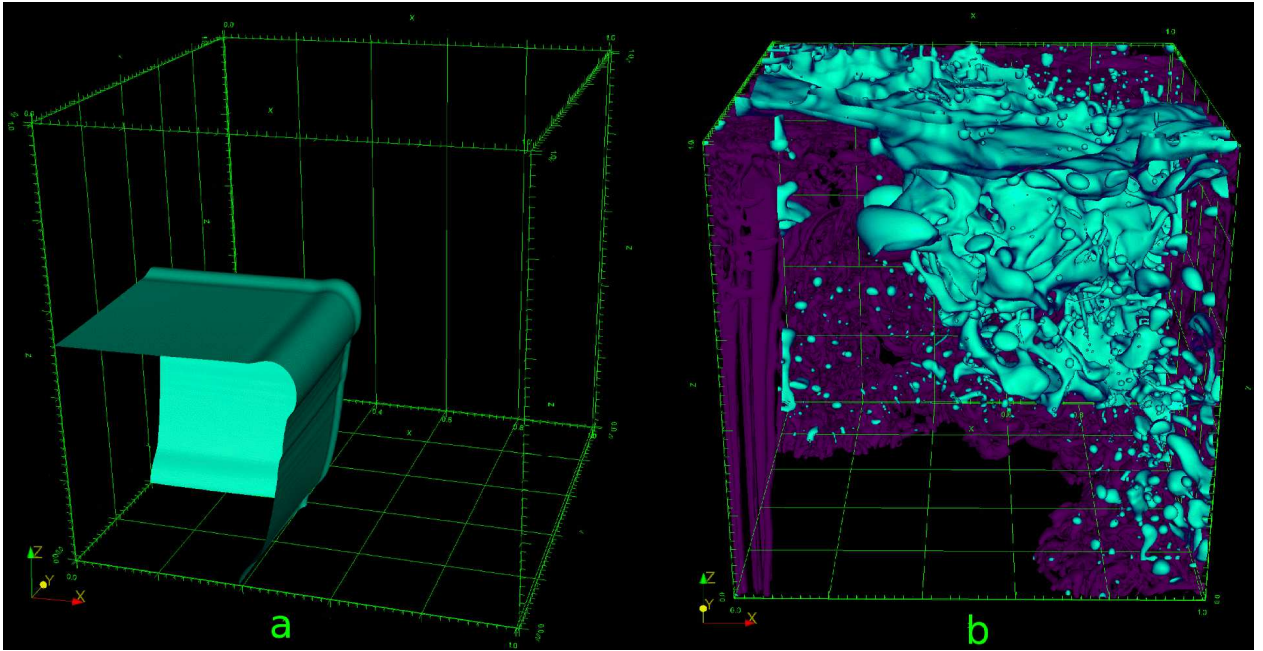}
  \caption{Separation of phases: macroscopic evolution of the flow. Simulation using the $256^3$ grid; (\textbf{a}) interfacial surface visible in initial stages of the flow at $t=0.46$s --  with entirety of the light phase (``oil'') still in the vicinity of the $(0,0,0)$ point; (\textbf{b}) fully developed flow at $t=6.25$s: interface visible with the $\lambda_2=-0.001$ isosurface shown in dark violet. }\label{sepfig5}
\end{figure}

In the following tests we will focus our attention on a ''separation of phase'' test-case, popularized mainly by the efforts of St\'ephane Vincent \cite{vincent2015,vincent_ASME}. The premise of this case is simple -- it features a cube-shaped domain $D$ of size $H$ in which we place two immiscible fluids of different densities and viscosities in the presence of gravity. What sets this apart from well-known Rayleigh-Taylor instability is that the fluids are not directly superimposed; instead, the lighter of them (''oil'', the other one being ``water'') occupies an octant nearest coordinate origin (Figure \ref{sepfig5}a). As a result, not only are fluids accelerated towards eachother (as in Rayleigh-Taylor instability) which occurs at the top wall of the light-fluid cube: additional shearing also takes place as the ``oil'' is lifted by buoyancy (see e.g. Fig. 14 in \cite{aniszewski2014caf}). The latter phenomenon is akin to Kelvin-Helmholtz instability. Macroscopic flow character depends on its Reynolds number. Examples are given in \cite{aniszewskiJCP} and \cite{vincent2015} which present behaviours ranging from that of a laminar bubble rise at $Re\approx 700, We \approx 11$ to a turbulent flow with strong interface fragmentation at $Re\approx 2\cdot 10^5, We\approx 10^2.$ Following Reynolds number definition is used\cite{vincent}:

\begin{equation}\label{sep1}
  Re=\frac{\rho_oHU_g}{\mu_o}=\frac{\rho_oH\left(\frac{\rho_o-\rho_w}{\rho_o}\sqrt{\frac{Hg}{2}}\right)}{\mu_o},
\end{equation}

where $o$ corresponds to lighter fluid (``oil'') and $w$ to the heavier one (``water''). The $U_g$ is a gravity-based velocity estimation, appearing also in the Weber number definition:

\begin{equation}\label{sep2}
  We=\frac{\rho_oHU_g^2}{\sigma}.
\end{equation}

The flow is of anisotropic character, with most of fluid acceleration taking place along domain diagonal, even for high-Re cases. Even as the ``oil'' undergoes fragmentation, the initial fluids volume ratio and asymmetry cause most light fluid parcels to be confined closer to two of domain's corners (see Figure \ref{sepfig5}b). This is different from a fully-developped Rayleigh-Taylor instability, in which the fragmentation phase is more isotropic\footnote{See Fig. 25 in \cite{aniszewski2014caf}.}. Regardless of its parameters, the flow concludes in a sloshing phase, after which fluid deplacement finishes with the light fluid occupying upper slice of the domain. Ideally, if no traced volume is lost, this slice should be $H/8$ high.

In all simulations presented in this section, a turbulent configuration has been chosen. The parameters are thus $\rho_o=900$ (kg/m$^3$), $\rho_w=1000,$ for densities, viscosity  $\mu_o=0.1$ (Pa$\cdot$s) with $\mu_w=0.001$ and surface tension coefficient $\sigma=0.45$ (N$/$m) with resulting $Re\approx 2\cdot 10^5.$  Example simulation is presented in Figure \ref{sepfig5}: the vortical structures developping around the interface are also visible there, visualized using Jeong's $\lambda_2$ criterion \cite{haller2005}. The figure showcases some of the characteristic features of this test-case, such as the coexistence of large- and small-scale interfacial formations. The bulk fluid visible in Fig. \ref{sepfig5}a is at one point largerly broken into droplet groups visible in Fig. \ref{sepfig5}b which in turn generate vortical ``trails''. Thus, the ability of the computational method to resolve the challenge of interface tracking is essential in this case as it generates also the enstrophy in both phases and regulates its transfer. Below, we will discuss if and how the ADM-$\tau$ model influences that process.

\subsubsection{ADM criteria comparison}

\begin{figure}[ht!]
  \centering
  \includegraphics[width=0.5\textwidth]{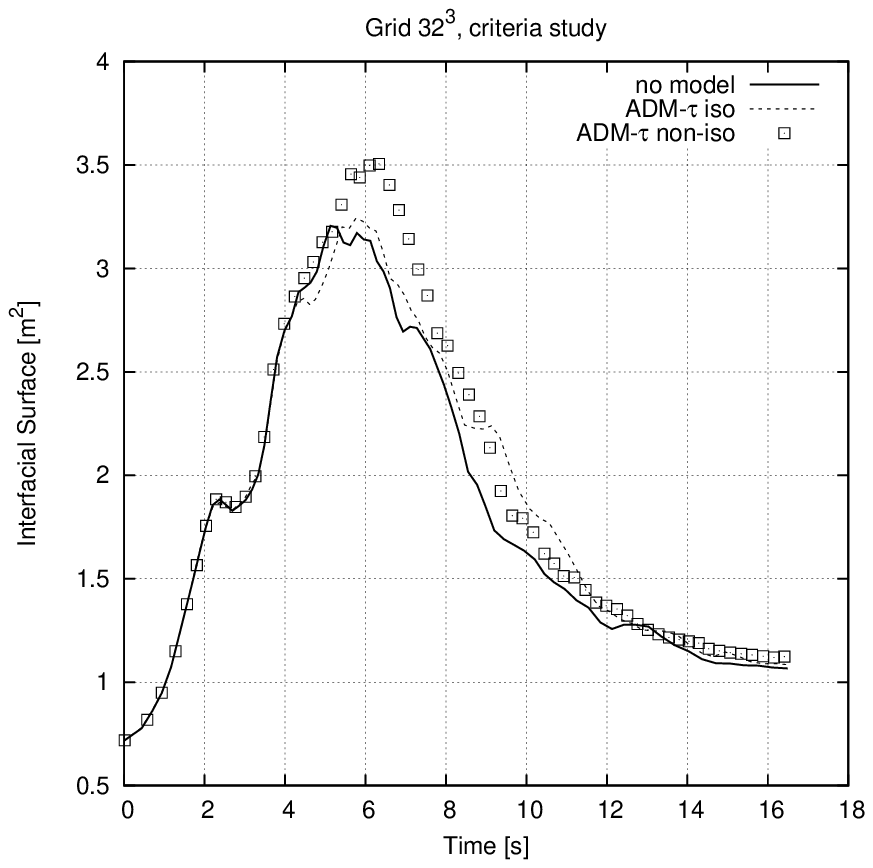}\includegraphics[width=0.5\textwidth]{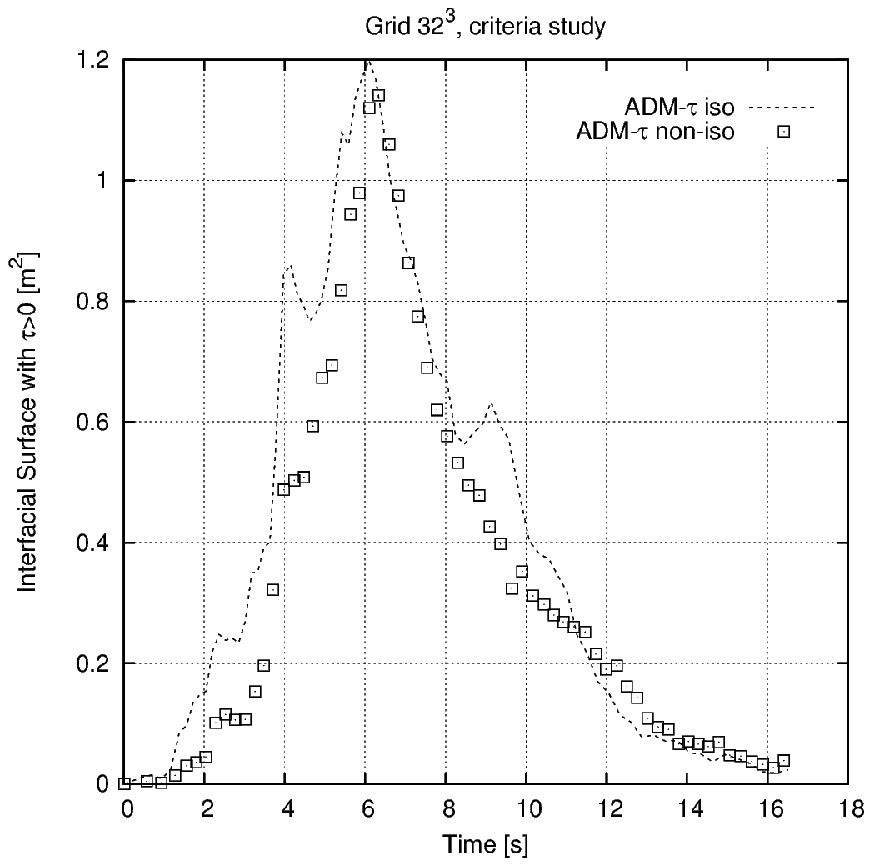}
  \caption{Phase Separation: (\textbf{a}) Interfacial surface resolved using the $32^3$ grid (left); (\textbf{b}) area of interface with $\trnn$ activated by the criteria (right).}\label{sepfig1}
\end{figure}

We continue our analysis of the newly proposed ADM-$\tau$ model variant by providing a brief study of the choice between the isotropic and non-isotropic criteria described in Section \ref{crit_sect}. The non-isotropic criterion is initialized  via (\ref{crit52}) with coefficients $c^{1,2}_\kappa=1,\epsilon=\frac{\kappa_{max}}{5.33}.$ In Figure \ref{sepfig1}a we can see the temporal evolution of the resolved interfacial surface \[S_{int}=\sum\limits_{i,j,k}S(\phi_{ijk},t)\] for the phase separation simulation\footnote{The $S(\phi_{ijk})$ is computed in individual cells by weighing $\phi$ values in the surrounding stencil.}. A coarse, uniform  grid of $32^3$ points was applied.  Using this sparse grid we are able to resolve less that $3.25m^2$ of area in the simulation with no model (peak of the continuous line). This value corresponds to time value of $t\approx 6$s, corresponding to the moment of maximum interface fragmentation: using a $64^3$ grid we obtain $S_{int}\approx 6m^2$ thus the interfacial geometry is definitely under-resolved in Fig. \ref{sepfig1}a.

Yet once ADM-$\tau$ is applied, the value increases by seven percent with the non-isotropic criterion. Using the isotropic one, the growth is significantly smaller. The $S_{int}$ increase itself is a desired effect as it represents the accuracy growth of this macroscopic variable, thus consituting a succesful test in the \textit{a posteriori} manner. This topic will be continued below.

Meanwhile, to interpret the difference in  $S_{int}$ yields of two considered ADM-$\tau$ criteria,  let us focus our attention on the accompanying Figure \ref{sepfig1}b. It displays the cummultiative area of the interface to which ADM-$\tau$ was non-trivially applied: that is

\begin{equation}\label{taus}
  S_{int}^\tau=\sum\limits_{i,j,k}S(\phi_{ijk},t)\cdot H\left(|\trnn| \right),
\end{equation}

where $H()$ is the Heaviside operator. Thus in Figure \ref{sepfig1}b, we see the actual surface extent with nonzero $\trnn$: it peaks at roughly $35\%$ of total interface area. Looking closer, we observe that the non-isotropic criterion actually makes the model affect  \textit{less} computational cells in total, which is intended due to having applied (\ref{crit12}). In other words, enabling the ADM-$\tau$ model to differentiate between various geometry configurations makes it more sensitive to actual interface topology, and act in more appropriate configurations. For example, relatively thin ligaments connected at both sides would have been acted upon all their surfaces by the isotropic ADM-$\tau$, while the newly presented version will act only at their selected sections, such as necks. This will accelerate structure's  breakup, procuring an effect consistent with Fig. \ref{sepfig1}a. 

\begin{figure}[ht!]
  \centering
  \includegraphics[width=\textwidth]{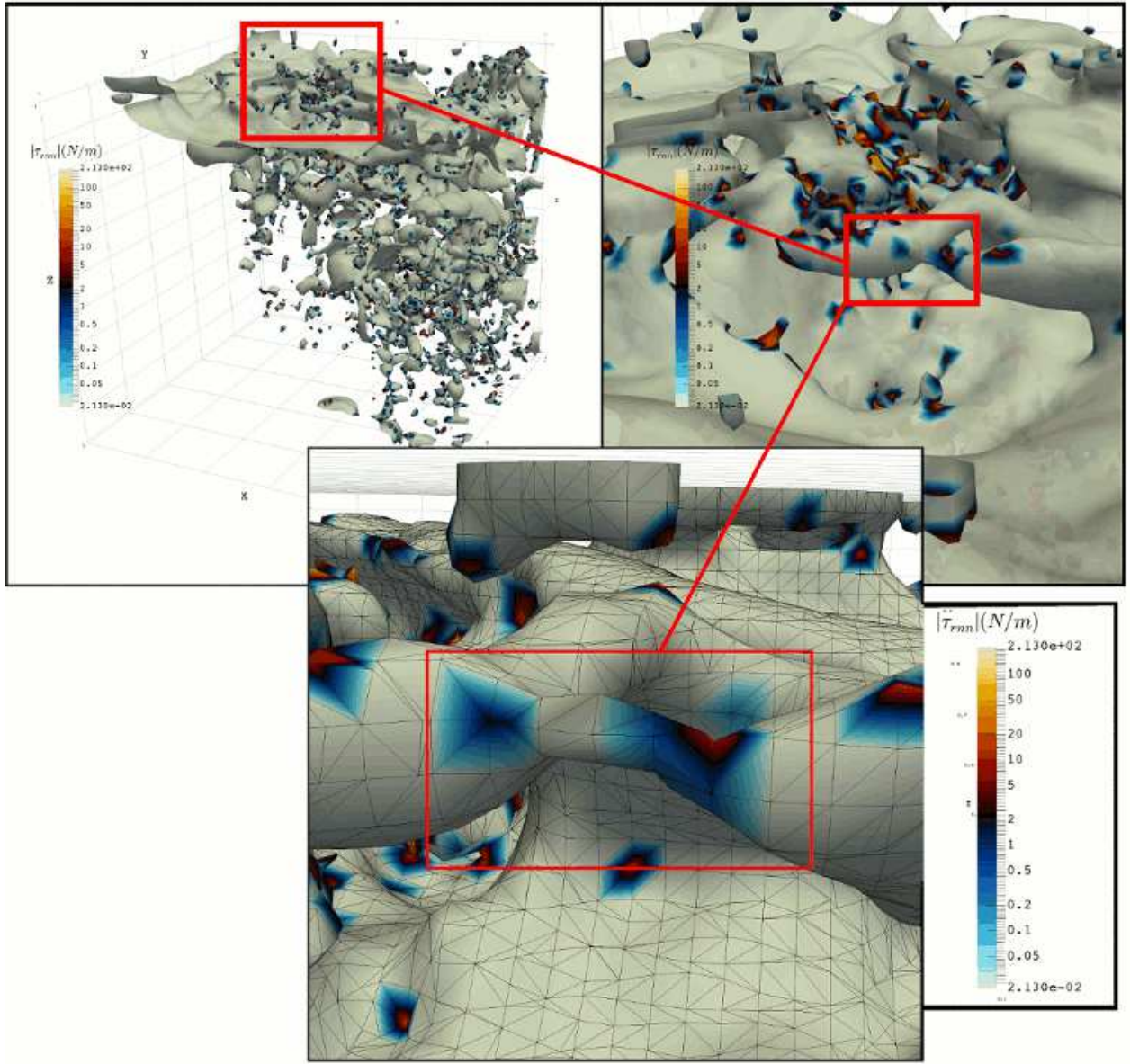}
  \caption{Phase separation simulation using the new, anisotropic version of ADM-$\tau$ model and a $128^3$ grid. The situation depicted corresponds to $t=7.07$s; three consecutive camera positions are shown for the same $t$ value. Clockwise from top-left: the entire interface visible in domain with a subsection in red containing a breaking-up ligament; the ligament visible with values of $|\tau_{rnn}|$ shown in color (logarithmic scale); the adjusted view of the said ligament drawn in coloured wireframe. (Note that the triangular wireframe structure is produced by the visualization software, and does not correspond to the actual solver discretization.)}\label{giant_separation1}
\end{figure}

The structure breakup mechanism accelerated by the anisotropic ADM-$\tau$ can be further shown in actual simulation, provided we can plot the $\trnn$ values produced by the model. Such an investigation is presented in Figure \ref{giant_separation1}. The simulation shown uses a $128^3$ hence the degree to which the model is active is smaller than on coarser grids. However, due to omnipresence of fragmentating structures, it is relatively easy to find examples of the breakup mechanism discussed. In Fig. \ref{giant_separation1}, top-left image presents whole simulation domain in which a fragmented interface is shown at $t=7.07$s.  This time value corresponds to a moment at which the number of cells with non-zero $\trnn$ is close to maximum (see Figure \ref{sepfig4}). Most of the interface however (coloured in Fig.
\ref{giant_separation1}with the $\trnn$ magnitude)  has the tensor value zero, due to the fact that the curvature is well resolved there.  (Note that in Fig. \ref{giant_separation1} the subgrid surface tension force vector arrows are not visible. Only their local magnitudes have been used to color the interface to increase readability.) We see that whenever the model is active, the magnitude of the computed tensor is mostly below $100$N/m which is consistent with the facts that $\sigma=0.45$ and the resolved curvature maximum value of $\max\kappa\approx161m^{-1}.$ Subsequent magnifications shown in Figure \ref{giant_separation1} focus on a ligament with non-zero $\trnn$ values at its ends, as discussed previously. This structure is on a brink of being resolved, and is characterized by large mean curvature, so the isotropic model version would produce non-zero tensor magnitudes on its whole surface, while the non-isotropic version does not affect part of its cells. This accelerates the breakup of this structure. 

\begin{figure}[ht!]
  \centering
  \includegraphics[width=0.8\textwidth]{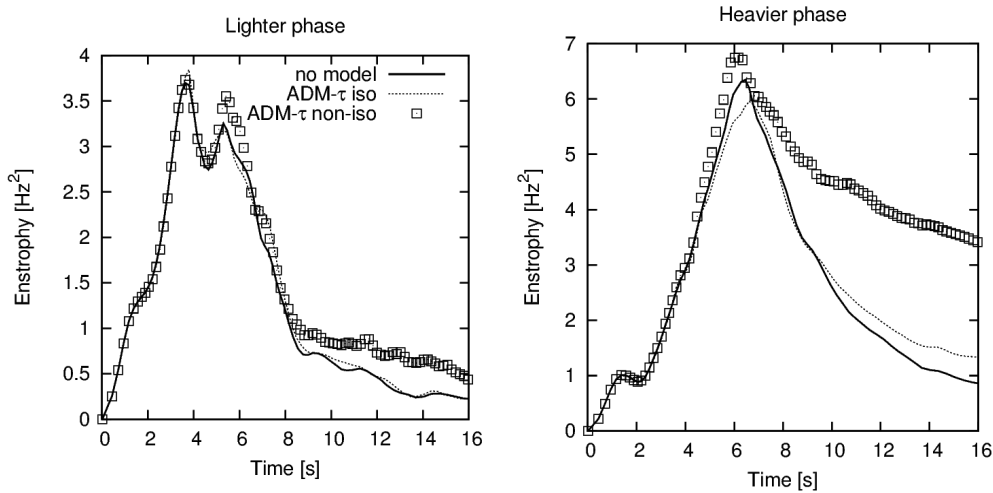}
  \caption{Temporal evolution of the enstrophy integral for the phase-separation simulation using a $32^3$ grid. (\textbf{a}) Lighter phase (``oil'') on the left-hand side, (\textbf{b}) heavier phase (``water'') on the right.}\label{sepfig2}
\end{figure}

As was shown in Figure\ref{sepfig1}, the peak value of the interfacial area $S_{int}$ occur at $t\approx 6$s, which is consistent with the estrophy peaks in both phases. Temporal evolution of the total enstrophy in both phases is shown in Figure \ref{sepfig2}. For this calculation, Volume of Fluid fraction function $C$ is used, and so for example, the value visible in Fig. \ref{sepfig2} for the lighter phase is:

\begin{equation}\label{ensdef}
  \Omega_2=\int\limits_DC\mathbf{\omega}^2dV
\end{equation}

with $\mathbf{\omega}$ denoting vorticity. Note that to obtain $\Omega_1$ it is sufficient to replace $C$ with $(1-C)$ in (\ref{ensdef}). Analyzing Figure \ref{sepfig2} we note that the peak of the ``water'' occurs after the one for ``oil''. In the lighter phase, the peak is dual: its first sub-peak corresponds to $t\approx 3.9$s when the large interfacial formation (bulk oil) is maximally stretched and its fragmentation begins. The latter, closer to $t\approx 6s,$ occurs when that fragmentation is maximum. This is when most kinetic energy is transported to the heavier phase -- as it can be seen in the righ-hand side image in Fig. \ref{sepfig2} -- and the vortical structures develop therein.

As one can infer from Fig. \ref{sepfig1}a, non-isotropic criterion yields the most interfacial area. It is the combined area of the fragmented interfacial formations (small ``oil'' droplets) whose number is greater if the ADM-$\tau$ non-isotropic is used. As those small droplets contribute to the vorticity and enstrophy of the ``water'' phase by exchanging their kinetic energy with it, the heavier phase enstrophy level visible in Figure \ref{sepfig2} is far higher when non-isotropic criterion is used. The same could be said with respect to the heavier phase, as evidently the non-isotropic model curve dominates at the second sub-peak, however it is the isotropic criterion curve rising highest in the first. This could hint at a different inter-phasal enstrophy exchange mechanism when the two criteria are used -- even more so as the heavier phase enstrophy peak in Figure \ref{sepfig2} is actually decreased with the isotropic criterion.

\begin{figure}[ht!]
  \centering
  \includegraphics[width=\textwidth]{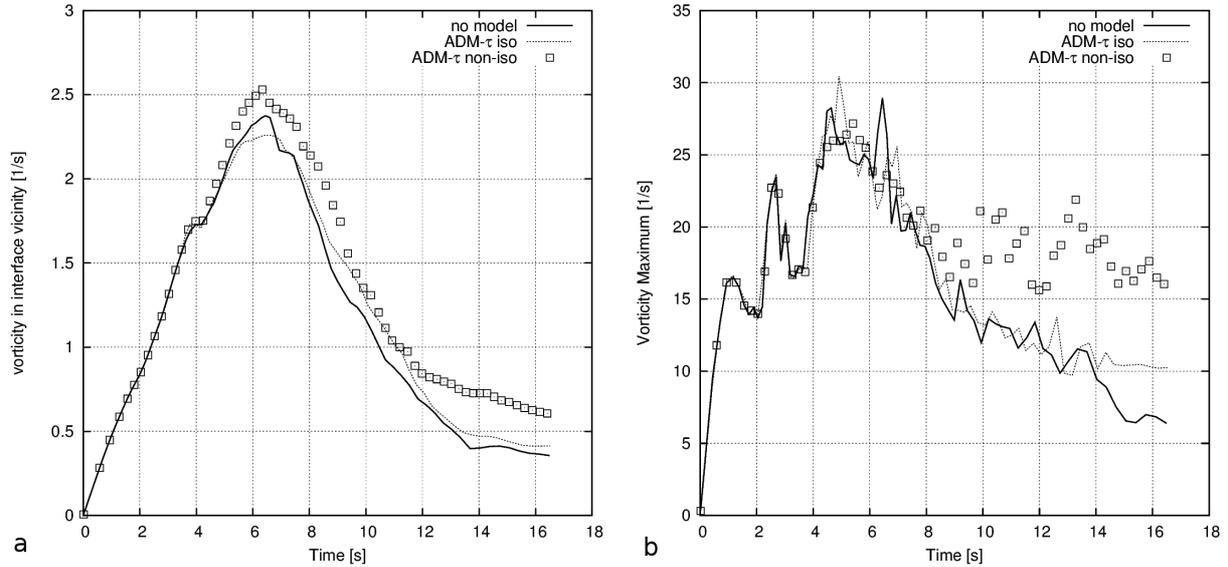}
  \caption{(\textbf{a}) Temporal evolution of the vorticity in interface vicinity (\ref{vsumeq}) for the phase separation simulation on a $32^3$ grid. (\textbf{b}) Global vorticity maxima found for the same simulation. }\label{sepfig3}
\end{figure}

Additional confirmation of the slightly more efficient kinetic energy transport for the non-isotropic criterion is found inspecting the values of vorticity itself: such information is presented in Figure \ref{sepfig3}. Figure \ref{sepfig3}a contains the plot of vorticity collected in interface vicinity defined as follows. Recalling that level-set distance function $\phi$ is zero at the interface: \[ \phi(\xb,t)=0 \Leftrightarrow x\in\Gamma,\] the \textit{discrete} neighbourhood $B$ of $\Gamma$ with radius $r$ can be defined as:

\begin{equation}\label{neigh}
  B_r(\Gamma)=\left\{ (i,j,k)\colon \phi_{ijk}(t)<r\right\}.
\end{equation}

Using the $\phi$ interval (\ref{neigh}), we can now collect scalar quantities in the spatial subset thus delineated. And so, for vorticity:

\begin{equation}\label{vsumeq}
  \sum\limits_{i,j,k\in B_{4\Delta x}(\Gamma)}|\omega_{ijk}(t)|
\end{equation}

means we collect vorticity over  all grid points for which level-set is smaller than a limiting value. \footnote{The definition is purposefully made grid-dependent to balance the increasing amount of resolved $\omega$ when decreasing grid size. However (\ref{vsumeq}) is not applied in this section in convergence context. }

Figure \ref{sepfig3}b displays the temporal evolution of $\max\limits_D|\omega|.$ As we can see, again the non-isotropic criterion produces dominant value in the post-breakup flow stages. Interestingly, the isotropic version produces a peak value of vorticity maximum, but the overall character of the dashed curve in Fig. \ref{sepfig3}b is similar to  that for which no model was used. Meanwhile, non-isotropic criterion both produces higher vorticity level in the interface vicinity, and produces dominant maxima in post-breakup phase: analysis of Figure \ref{sepfig2}b suggests that those maximum values occur in the ``water'' phase, as the non-isotropic criterion causes much more vortical structures to appear there.

This concludes the study of the criterion choice for the ADM-$\tau.$ For the reminder of the section, only the results obtained using the non-isotropic-criterion will be presented\footnote{Additional comparative material, this time  concerning ADM-$\tau$ model variants A and B can be found in \ref{apx}.}. The criterion will be used with equal coefficients $c^{1,2}_\kappa=1$ which, is discussed in Section \ref{crit_sect} ammounts to 'ballanced' anisotropy. Preliminary tests with non-equal, fixed  values have been performed using the same setup with promising results. However, if a  procedure with any physical significance is envisaged  -- such as preference of isotropy for disperse bubbly flows, or preference of anisotropy for stretched ligaments in non-newtonian fluids -- dynamic setting of $c^i_\kappa$ is needed. Such results will be presented in future publications.

\subsubsection{Model convergence and \textit{''a posteriori''} performance}

\begin{figure}[ht!]
  \centering
  \includegraphics[width=0.6\textwidth]{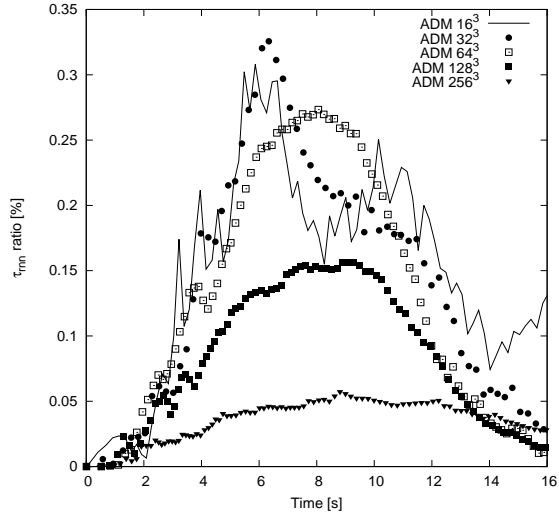}
  \caption{The percentage of interfacial surface during the separation of phases simulation with non-zero $\trnn$ for five grid sizes.}\label{sepfig4}
\end{figure}

The convergence of the model is portrayed by the decreasing values of surface area with non-zero values of $\trnn$ (analogous to Figures \ref{osc_amp1}b and \ref{sepfig1}b), as shown in Figure \ref{sepfig4}: we see that using the coarser grid, a third of interfacial area has non-zero tensor values for $t\approx 6$s.  Subsequently, ratios diminish down to five percent for the finest grid.

While the general convergence is unquestionable for grids denser than $32^3,$ it is roughly stagnant for the first two grids, which is likely due to the fact that the $16^3$ is far too coarse to enable proper traced volume conservation in such a demanding test-case (approximately $20\%$ volume lost).

On the other hand it is noticeable that the ratio pictured in Fig. \ref{sepfig4} has apparent peaks at $t\approx 6$s for coarser grids and at $t\approx 8$s otherwise, suggesting that once the grid is changed from $32^3$ to $64^3,$ a specific scale of  interfacial formations appear (i.e. becomes resolvable) that influences the character of simulated flow. Also these formations are characterized by nearly-resolved curvatures, thus enabling $\trnn$ contribution to increase. Indeed, once the bulk light fluid stretches and breaks up at $6<t<8$, numerous horizontal ligaments are formed which only appear using grids of  $64^3$ points and above.  Being barely resolved, they are characterized by nonzero $\trnn$ values. Additionally, they contribute to the production of vortical structures, which is  reflected in the flow kinetics, as shown below (see also Figure \ref{giant2}). %[BTW I HAVE 64**3 ANIMATION - could be POSSIBLE IN JCP?]

As a way to assess the performance of the model in a real physical flow context of a non-resolved simulation, we propose the study of the enstrophy integral (\ref{ensdef}). While other macroscopic variables, such as total kinetic energy or the aforementioned $S_{int}$ may be roughly converged for this flow configuration with presented grids, the vortical structures are still found non-resolved for grids up to $1024^3$ and higher \cite{vincent2015}. If $N$ stands for the number of uniform mesh elements in each direction, the sum of resolved vorticity rises as $O(N)$ (not shown). The derived enstrophy exhibits similar behaviour, hinting at a hypothesis that the energy transfer in this flow involves a cascade of scales similar to that of other turbulent flows. This is why the enstrophy is an useful quantity when it comes to testing LES closures in \textit{a posteriori} manner.

\begin{figure}[ht!]
  \centering
  \includegraphics[width=0.5\textwidth]{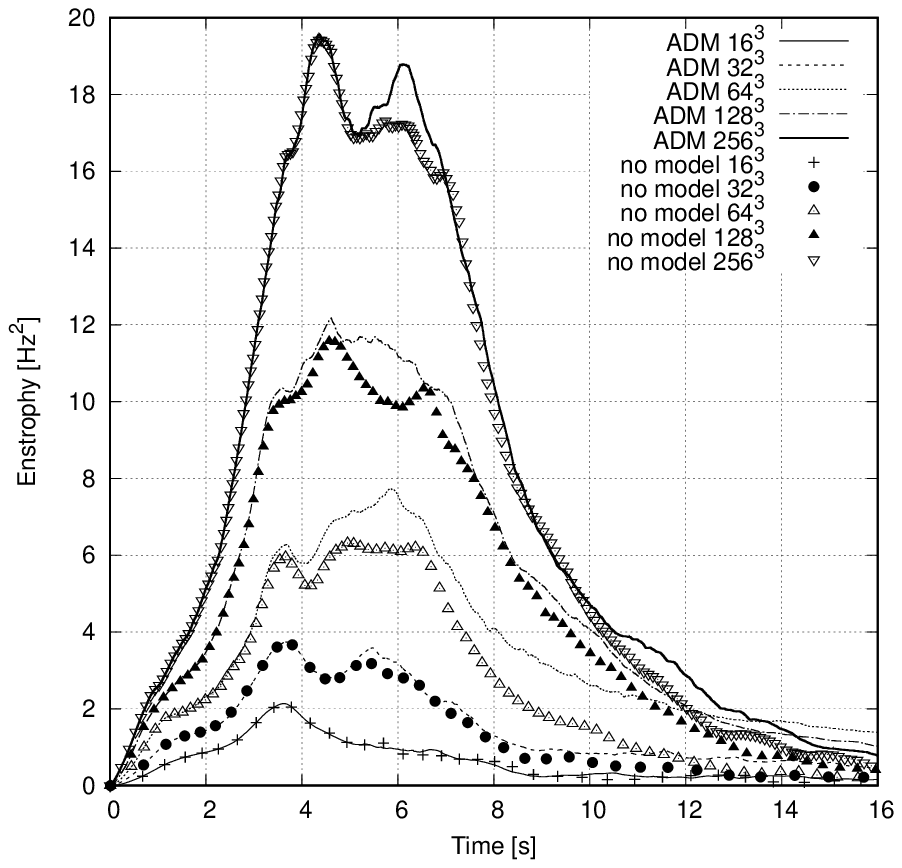}\includegraphics[width=0.5\textwidth]{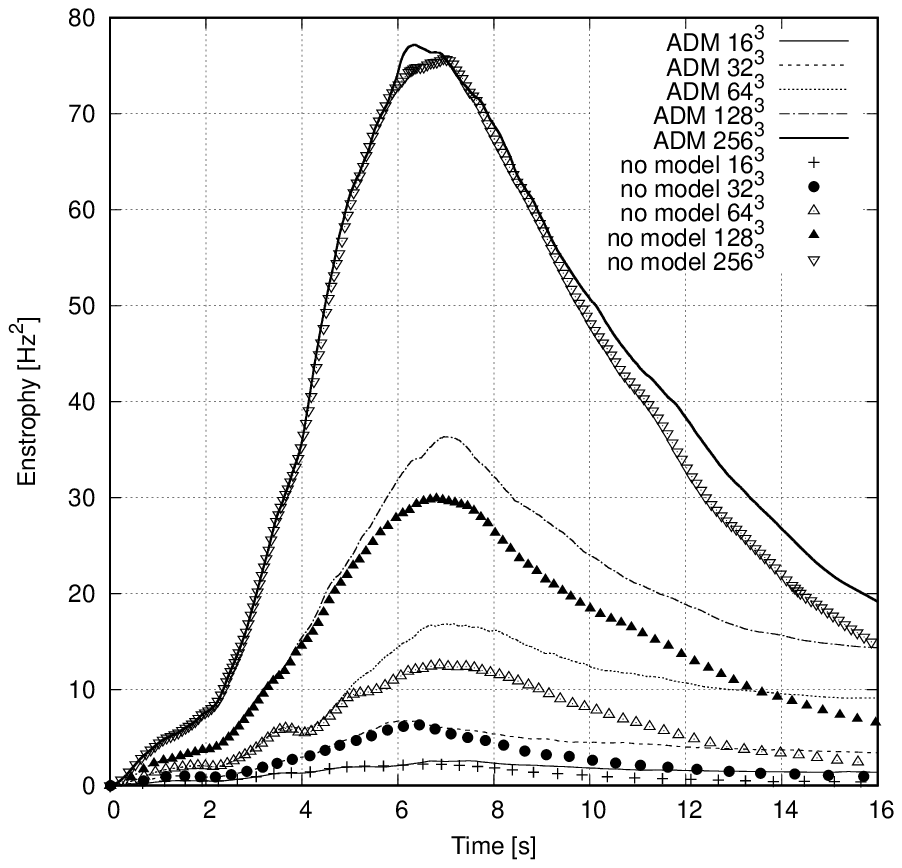}
  \caption{The enstrophy curves for the phase separation test-case using varying grid size. (\textbf{a}) For the lighter (``oil'') phase, (\textbf{b}) the heavier (``water'') phase.}\label{sepfig42}
\end{figure}

The temporal evolution of (\ref{ensdef}) is presented for both phases in Figure \ref{sepfig42}. For each of the grids, the curve plotted with points illustrates the simulation without the tested model, while the lines represent the ADM$-\tau$ performance. Due to the continuous rise in resolved enstrophy, the curves for each grid form a pair distinct from the other, hence enabling us to observe both model performance and simulation convergence in one graph.

Inspecting the plots in Fig. \ref{sepfig42}a which illustrates the lighter phase, we first observe the dual character of enstrophy peaks. The physical mechanism behind it is that around $t\approx4$s (the first sub-peak) the bulk lighter fluid reaches the domain corner opposite to its initial position; in other words the ``oil'' travels from point $(0,0,0)$ to $(1,1,1).$ The subsequent peak at $t\approx 6$s occurs once the large scale  vortical swirl\footnote{See Figure 15 in \cite{aniszewskiJCP}.} transports the bulk liquid so that the interface reaches the $(1,0,0)$ point -- this is when most interface fragmentation occurs. It is also when the enstrophy peak is formed in the second phase (Fig. \ref{sepfig42}b).

Enstrophy is an important statistic describing the flow as it provides the information on accuracy, to which the vortical structures are resolved \cite{vincent_ilass,pope}. Plots in Figure \ref{sepfig42} show  the increase in this accuracy due to ADM-$\tau$. In the lighter phase it is visible after the ``breakup'' stage ($t\approx 6$s) and into subsequent  ``emerging'' and ``sloshing'' stages. Actual peak values are given in Table~\ref{ens_tab}: the model-induced gain  is most visible using the $64^3$ grid. This is related to the ligament formations that appear at that grid resolution, as mentioned above in context of Figure \ref{sepfig4}.

Note however that when considering the light phase, only the instantaneous values at absolute enstrophy peaks  are visible in Table \ref{ens_tab}, which does not account for these peaks' dual character.  Thus e.g. for the $256^3$ grid (Fig.\ref{sepfig42}a: topmost curves: thick line/ white triangles) Table \ref{ens_tab} denotes  a mere $0.68\%$ enstrophy increase (as the values for $t=4.36$ are accounted for)  however almost a $10\%$ enstrophy increase is noted for the second sub-peak at $t=6.12$s.

\begin{table}
  \begin{center}
    \caption{Separation of phase: peak enstrophy values for simulations with and without the model.}
    \begin{tabular}{l|c|c|c|c}
      &  \multicolumn{2}{|c|}{No model} & \multicolumn{2}{c}{ADM-$\tau$} \\ \hline
      grid   &   oil      &   water     &   oil        &    water    \\ \hline
      $16^3$ &   $2.133$  &   $2.243$   &   $2.135$ $(+0.09\%)$    &    $2.591$ $(+15.5\%)$ \\
      $32^3$ &   $3.706$  &   $6.326$   &   $3.737$ $(0.8\%)$    &    $6.799$ $(7.47\%)$ \\
      $64^3$ &   $6.337$  &   $12.540$   &   $7.735$ $(22\%)$    &    $16.7$ $(33.1\%)$ \\
      $128^3$ &  $11.571$  &   $29.881$   &   $12.165$ $(5.13\%)$    &    $36.339$ $(21.6\%)$ \\
      $256^3$ &  $19.430$  &   $75.711$   &   $19.564$ $(0.68\%)$    &    $77.187$ $(1.87\%)$ \\ \hline

    \end{tabular}\label{ens_tab}
  \end{center}
\end{table}

Similar conclusions could be drawn from inspection of the enstrophy evolution in the heavy phase (``water'') seen in Fig. \ref{sepfig42}b. The striking $33\%$ and $22\%$ increases in enstrophy for the $64^3$ and $128^3$ grids respectively clearly demonstrate the model effectiveness. Favorable result is however obtained using all discretizations. Intrestingly, in the later flow stages ($t>10$s) the enstrophy resolved in the ADM-$\tau$ simulations can reach levels of the no-model simulation done with grid twice finer: this however seems to be the discharge of vortical structures created by ADM-$\tau$ in the ``breakup'' phase of $6<t<8$s.

It seems plausible that the transfer of $E_k$ from light to heavy fluid takes place mostly in that very phase, which would also help to explain why only a single peak is obtained when calculating (\ref{ensdef}) in ``water''\footnote{More discussion on the physical character of this flow can be found in \cite{aniszewskiJCP, vincent} and \cite{vincent2015}. }.

\begin{figure}[ht!]
  \centering
  \includegraphics[width=0.666\textwidth]{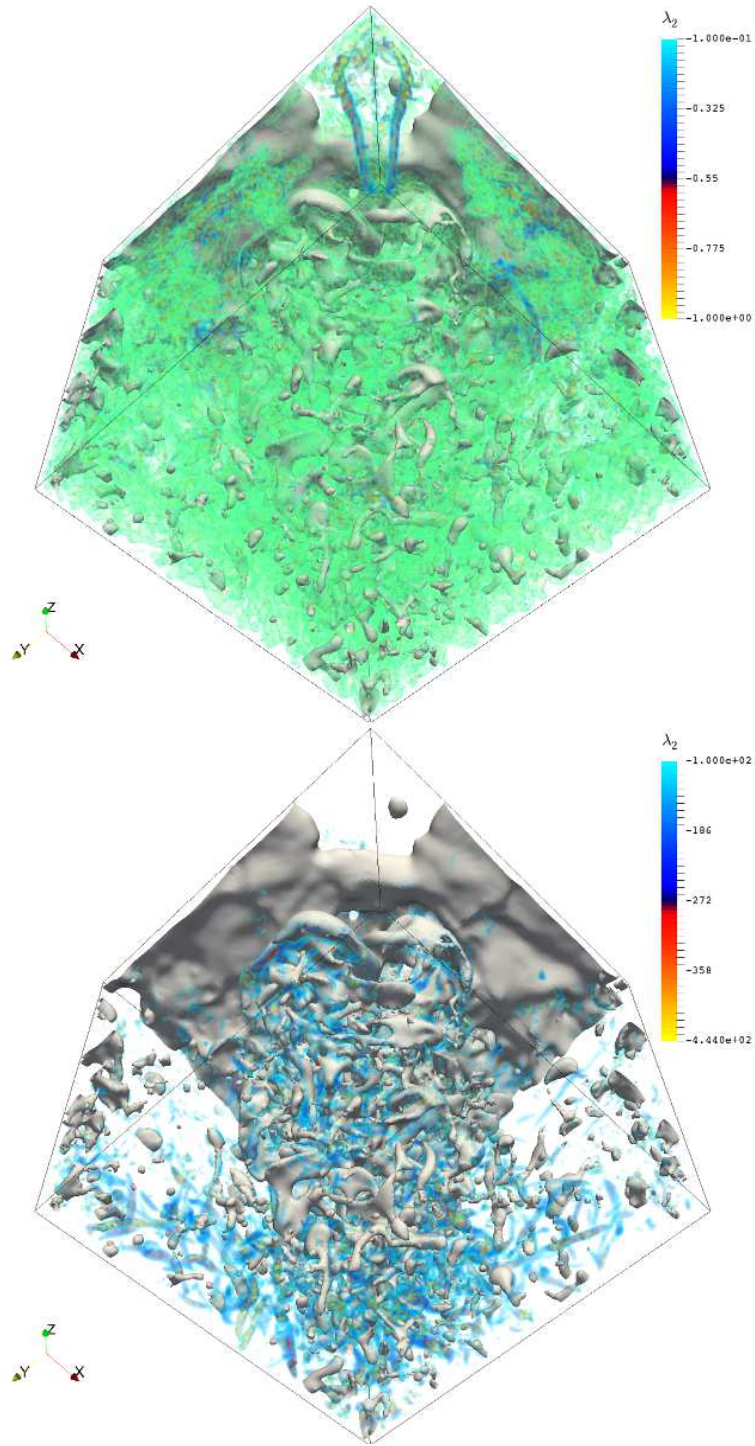}
  \caption{Visualization of vortical structures in phase separation simulation. The $\lambda_2$ criterion \cite{haller2005} scalar field has been presented using volumetric rendering: the ``liberal''  $\lambda_2\in\lb-1,-0.1\rb$ range in the top picture, the more strict range $\lambda_2\in\lb -440,-100\rb$ at the bottom. Interfacial $\phi=0$ surface shown in white in both pictures.}\label{giant2}
\end{figure}

As has been said, it is the heavy phase (``water'') in which the enstrophy increase due to ADM-$\tau$ is most significant. Therefore it is interesting to illustrate the vortical structures in this phase, especially in post-breakup stage of the flow -- so for $t>7$s, after the enstrophy peak value occurs. We include a visualization of the vortical structures in Figure \ref{giant2} for a simulation that used the $128^3$ grid and ADM-$\tau$ model. Two images are included, showcasing the interface (white $\phi=0$ isosurface) and the volumetric rendering of the $\lambda_2$ vortex criterion \cite{haller2005}. The view is from below the domain. Since the lighter fluid at this flow stage has already floated up, it occupies mostly the vicinity of the top wall, hence mostly heavier phase is visible here.

In the top picture, a less requiring (``liberal'' \cite{ll2011}) $\lambda_2\in\lb-1,-0.1\rb$ values are rendered. This reveals entire domain full of vortical structures, making the image hardly readable, with nearly all interfacial formations wreathed in vortices. However, two distinct vortex tubes are still visible in ``water'' in the top corner. These are vortex trails that follow the bulk ``oil'' as it is lifted by buoyancy. We see that even if the light fluid itself is long gone, the structures haven't dissipated yet.

Subsequently, the bottom picture of Figure \ref{giant2} showcases more requiring $\lambda_2\in\lb -440,-100\rb$ range (we denote that $\min_{t=7}(\lambda_2)\approx 3000 $). Those are  more energetic vortical structures in the ``water'' phase which, not surprisingly, are grouped mostly in the breakup region. Visible are series of horizontal vortices that have been ejected during breakup of the lighter fluid ``curtain'' after it has reached the $(1,0,0)$ point  which, as mentioned above, takes place just before the second enstrophy peak in Fig. \ref{sepfig42}b occurs. They gradually recede from the breakup region towards the lower domain edges, after which they are dissipated. However as confirmed in Figure \ref{sepfig42}b, their number is significantly greater once  ADM-$\tau$ model is used.

\section{Conclusions and Discussion}

We have presented a new version of the ADM-$\tau$ model for calculation of the sub-grid surface tension contribution to the surface tension force in Large Eddy Simulation of two-phase flow. The paper builds upon the previous publication -- that first introduced the method -- namely a simpler variant is proposed here. In this newly presented ADM-$\tau$ variant (``A'') direct deconvolution of phase indicator function (and thus interface normals) and curvature is performed, which allows us to calculate the subgrid term using its definition without any additional assumptions -- except certain model constants for flexibility. This is considerably simpler than any other method of $\trnn$ calculation, including variant ``B'' of the same ADM-$\tau$ method, as this time no additional advection step is required. Also, variant ''A'' is adapted to Level Set method and no longer requires Volume of Fluid function, broadening the scope of ADM-$\tau$ applicability\footnote{It has to be admitted however, that having reconstructed the velocity in the original ``B'' variant can be considered an asset: not only does it allow to perform the ADM-$\tau$ algorithm, but also address other modelling such as $\tau_{luu}$ or (presumably) $\tau_{\mu D}$ terms.}. The model amplifies the influence of small-scale (sub-filter) curvature, and constitutes a step towards closure of filtered Navier-Stokes equations (\ref{eq10}). 

 Note that traditionally in LES, the influence of the $\ub^{SGS}$ is in most approaches only modelled. The original ADM method of Stolz \& Adams  aims at reconstructing this subgrid velocity, and so does  the ADM-$\tau$ "B" variant published in \cite{aniszewskiJCP}. However $\ub^*$ obtained therein is not strictly equal to $\ub=\udash+\ub^{SGS}$ as the $G^{-1}$ operator is approximated (hence the method's name). Still, $\ustar$ is used to interact with the interface, whose resulting displacement is used to derive the $\trnn$ force. The deconvoluted velocity  is however not applied for modelling of (\ref{eq9}) and closing the $\tau_{u\phi}$ term and the right-hand side of filtered phase-indicator function advection equation (\ref{eq10b}) is taken as zero. As such, ADM-$\tau$ model (in all variants) does not \textit{directly} initiate the interaction between subgrid velocity and interface. However, it does so \textit{indirectly}, by modifying the surface tension force depending on whether interfacial curvature is resolved or not. This indirect character is emphasized even stronger in the newly presented variant ''A'' of our method, as the reconstructed velocity $\ustar$ does not appear in this formulation. We believe that for full interface-turbulence interaction or, in other words, its interaction with subgrid velocity, the (\ref{eq9}) must be closed as well.

 It is hard to find a test case in which a DNS result could be closely replicated by this model. But we note that even a partial closure of (\ref{eq10}) by $\trnn$ yields definitive improvement in results. The proposed body of tests constitutes a convincing proof that ADM-$\tau$ is an applicable solution for the closure of sub-grid surface tension, and its inclusion has expected influence on performed simulations in the \textit{a posteriori} sense. In that vein, with ADM-$\tau$ we have e.g. managed to obtain accumulated enstrophy distributions in phase inversion problem corresponding to $50\%$ finer DNS grids without any added CPU cost. In other test, the non-resolved droplet oscillation period is shown to have its error decreased once ADM-$\tau$ is used.

Finally, in the paper we have presented several improvements over the original ADM-$\tau$ variant \cite{aniszewskiJCP}. The version presented herein is anisotropic, as it includes a criterion which, by taking into account the resolved principal interface curvatures, modifies $\trnn$ value accordingly. Tests indicate that in a phase-separation test case, in which surface-tension contribution is a significant factor controlling the complex breakup process, the anisotropic model vastly outperforms the previous, isotropic version. Even in its anisotropic formulation, however, ADM-$\tau$ remains by far the simplest existing subgrid surface tension calculation scheme, compared with other authors' work. It is also much less CPU-intensive than its original formulation (see \ref{apx}). Similarly, the paper shows that properly controlled deconvolution of $\phi$ could be used further, for example in an effort to model other terms in two-phase LES, such as the (\ref{eq5}) tensor containing viscosity  term that depends on the distance function. This also should be the part of our research that will ensue this paper. 

\appendix
\section{ADM-$\tau$ Variants Comparison}\label{apx}

This section presents additional material regarding the comparison of the ADM-$\tau$ variant ``A'' this paper introduces, and the ``B'' variant as presented in \cite{aniszewskiJCP}. For this task, full re-implementation of variant B has been performed within  Archer3D solver (as per Fig. \ref{tau_diagramB}), thus creating a choice of ADM-$\tau$ variant for the programmer within otherwise identical code. A series of phase separation simulations similar  to those of Section \ref{sephase_sect} has been performed, with the same flow parameters. The tests are accompanied by an example CPU-cost comparison.

In the tests below, variant A is presented as in Figure \ref{adm_varA}, i.e. with all stages and non-isotropic criterium discussed in section \ref{crit_sect}. Variant B is presented exactly as \cite{aniszewskiJCP}, depicted in Figure \ref{tau_diagramB}, which is motivated as follows.

The convergence criteria of variant A (stages III and IV) can be seen as versatile $\trnn$ post-processing: they could be entirely omitted from the variant A; it is just as well possible to include them in variant B.  However, including all possible  combinations  here  would lead to a very complex, multi-way comparison, by presenting each variant with or without criteria, which in turn could be isotropic or non-isotropic. This would fall outside the scope of this paper - whose purpose is introduction of variant A; as such it should be a subject of later research. Thus we settle upon presenting both variants only in their default configurations.

\subsection{Flow Characteristics}

\begin{figure*}[ht!]
  \centering
  \includegraphics[width=0.8\textwidth]{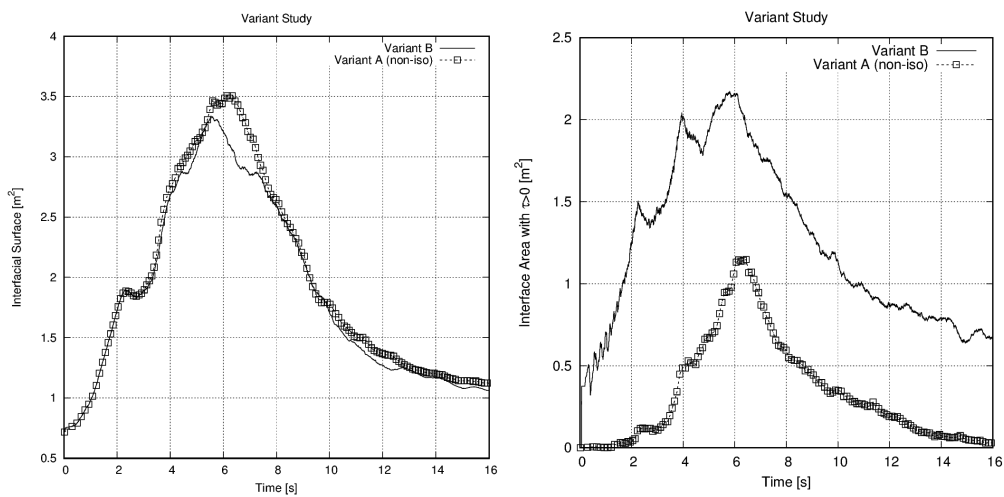}
  \caption{Resolved interface area (left) and interface area with $\trnn$ active (right) for phase separation simulations using ADM-$\tau$ variant A (squares) and B (continuous line). }\label{apx_int}
\end{figure*}

In Figure \ref{apx_int}, resolved interface area is shown for phase separation simulations using both variants. Clearly, simulation using variant A yields higher area (continuous line, note the curve is identical to Fig. \ref{sepfig1}a). Still, even the interface area computed via variant B is above levels obtained without the model (visible in Fig. \ref{sepfig1}a). Inspecting Figure \ref{apx_int}b, which presents interface area with non-zero $\trnn,$  we see that variant B yields nearly twice the value of variant A ($2.2$ over $1.1m^2$) meaning it has been activated far more often. This is obviously due to the action of A's convergence criteria. As mentioned in context of Figure \ref{giant_separation1}, this  more selective action of variant A, leading to sparser $\trnn$ distributions,  may procure higher traced mass fragmentation, which seems consistent with result in Fig. \ref{apx_int}.

\begin{figure*}[ht!]
  \centering
  \includegraphics[width=0.8\textwidth]{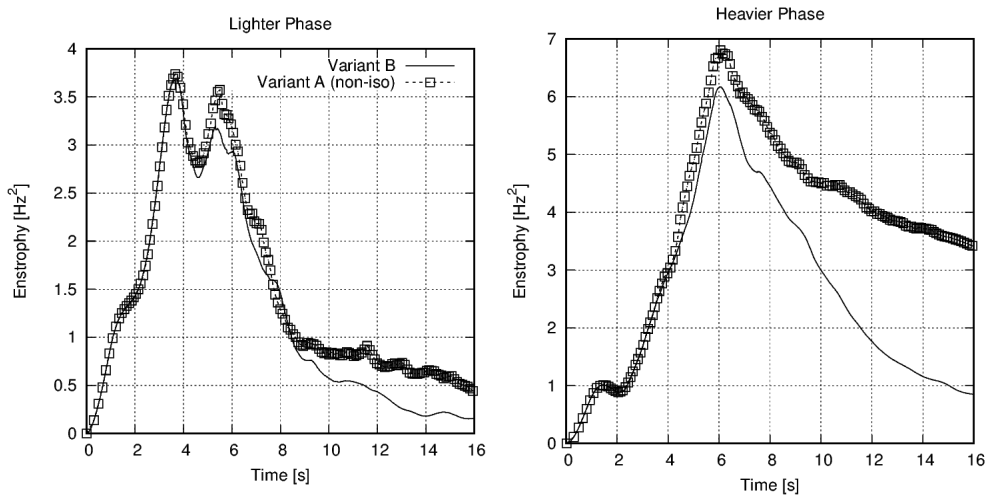}
  \caption{Enstrophy curves in both phases for phase separation simulations using ADM-$\tau$ variant A (squares) and B (continuous line). }\label{apx_ens32}
\end{figure*}

Figure \ref{apx_ens32} depicts the temporal evolution of enstrophy in both phases, thus recreating Fig. \ref{sepfig2} only this time with variants A and B compared. The latter produces a curve remarkably similar to the A's isotropic criterion curve in Fig. \ref{sepfig2}, i.e. enstrophy peaks in lighter phase (Fig. \ref{apx_ens32} left) are comparable, while in heavier phase variant B produces less enstrophy, especially in later flow stages. Interestingly, the peak of enstrophy in heavier phase using variant B is at $6.1$ Hz$^2,$ while variant A isotropic reaches $6.0.$ In any case, values obtained with non-isotropic variant A remain unparalleled in Fig. \ref{apx_ens32}.

\begin{figure*}[ht!]
  \centering
  \includegraphics[width=0.666\textwidth]{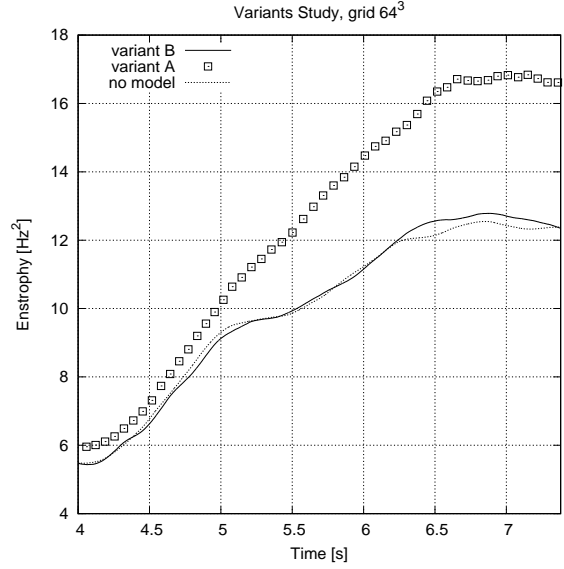}
  \caption{Phase separation simulation using $64^3$ grid: enstrophy peaks for both investigated ADM-$\tau$ variants and simulation w/o the model. }\label{apx_ens64}
\end{figure*}

We conclude the discussion of enstrophy evolution turning our attention to Figure \ref{apx_ens64}. In this  Figure, peak values of enstrophy are presented for simulations using the somewhat denser $64^3$ point grid. Heavier phase is considered, with the temporal range narrowed to $t\in\lb 4,7.5\rb$ allowing us to focus on the peak values. As previously, values produced by variant B (continuous line) fall below that of variant A (squares). However this time, due to the 'no model' curve plotted (dashed line) we are able to observe the increase in enstrophy due to variant B over the simulation without the model. There is thus an improvement in this flow characteristic once ADM-$\tau$ is used even in version presented in \cite{aniszewskiJCP} although not as pronounced as using the variant presented here.

Summarizing this subsection, we state that the inclusion of ADM-$\tau$ variant B \cite{aniszewskiJCP}, while seemingly beneficial for the weakly-resolved simulation -- as shown in Fig. \ref{apx_ens64} -- appears less effective than the variant A introduced in this paper. The same conclusions can be drawn inspecting results such as sum of kinetic energy or vorticity (not shown).

We continue the comparison by investigating the computational cost of both variants.

\subsection{CPU Cost}

Inspection of Figure \ref{adm_varA} shows clearly that $\trnn$ is obtained once stages I and II of variant A are completed. Hence these stages  themselves replace entire variant B presented in Figure \ref{tau_diagramB}. In terms of computational cost, stages I and II of variant A are obviously cheaper than variant B. The former only take deconvolution of two 3D arrays: $\kappa$ and $\phi;$ while the  latter not only comprises of deconvolution of \textit{three} 3D arrays (components of velocity) but \textit{additionally} a full  advection step, which in a CLSVOF framework is a multi-step operation \cite{menard} far outweighing deconvolutions.

\begin{figure*}[ht!]
  \centering
  \includegraphics[width=0.9\textwidth]{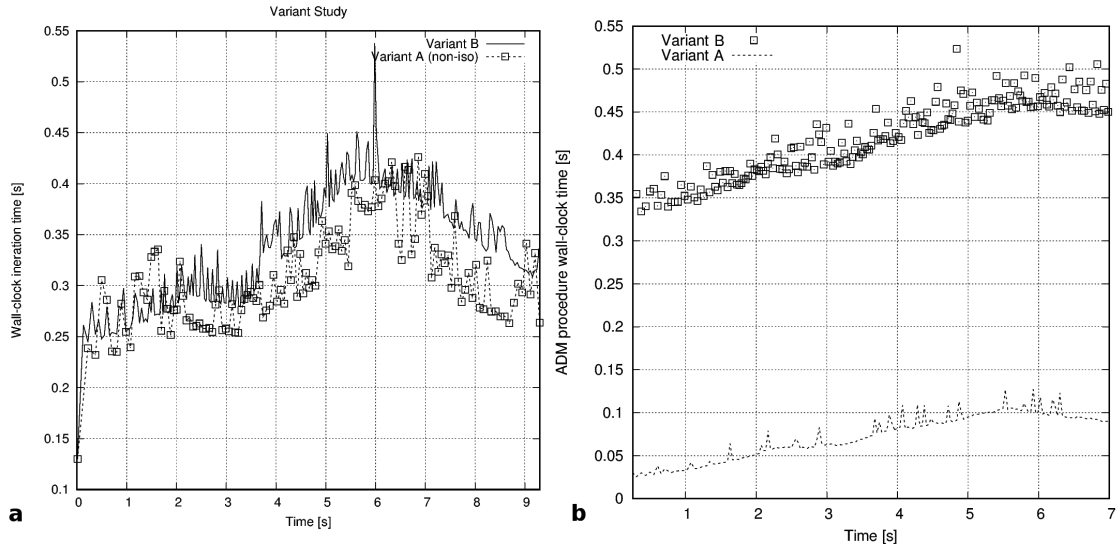}
  \caption{Phase separation test case, $32^3$ grid: (a) wall-clock time per complete solver iteration with ADM-$\tau$ model variants A and B; (b) wall-clock time due to ADM procedures only, for the same simulation.  }\label{apx_wallclock}
\end{figure*}

However, even with criteria applied (all stages in Fig. \ref{adm_varA}), in actual simulations variant A proves less CPU-intensive than variant B as shown in Figure \ref{apx_wallclock}. In this Figure, phase separation case of Section \ref{sephase_sect} is presented, calculated using variants A and B. The $x$ axis denotes physical flow time. Figure \ref{apx_wallclock}a  shows wall-clock iteration time (in seconds) measured in a single core\footnote{Intel i7 processor, Sandy Bridge architecture, 2.7Ghz.}, single thread calculation. We see that using $32^3$ grid, average iteration time is close to $0.3s$. Shorter duration of iterations using ADM-$\tau$ variant A is clearly visible, with average gain of order of $10\%,$ peaking at $20\%.$ This however is not conclusive, since full Archer3D iteration consists of standard projection step, of which subgrid modeling is merely a fraction. Also, as shown above, inclusion of chosen model variant for $\trnn$ influences flow character, so the result in Fig. \ref{apx_wallclock}a is subject to feedback mechanism. Thus, we have included in Figure \ref{apx_wallclock}b, the wall-clock time due to ADM procedures \textit{only}. This is obtained by proper bracketing inside the code using $\mathtt{mpi\char`_wtime()}$ procedures\footnote{MPI-Message Passing Interface, procedures in Fortran standard, see \url{http://www.mpi-forum.org/docs/mpi-3.1/mpi31-report.pdf}.}. Using also a single core, $32^3$ calculation, we see that -- once other solver operations no longer disturb the measurements --  computational cost of variant A turns out to be up to an order of magnitude lower than that of variant B. The smallest gain is denoted for $t\approx 5.5s$ at which variant A is still five times faster than variant B.

\section*{Acknowledgements}

Part of this work has been supported by the National Research Agency (ANR) program Investissements d'Avenir: ANR-10-LABEX-09-01 EMC3-TUVECO (Turbulence et Viscoélasticité dans les Ecoulements COmplexes) research project at CNRS UMR6614 CORIA Rouen. The calculations have in part been performed at CRIANN (Centre Régional Informatique et d'Applications Numériques de Normandie). Entire work has been performed using Debian GNU/Linux operating system; visualizations were done using  Paraview \cite{kenneth} and Gnuplot \cite{gnuplot}.

The author would like to express gratitude to T. M\'enard for his help concerning principal curvature calculations; as well as  to M. Marek and  A.A.S.  for reading the manuscript.  I would like to thank  S. Vincent, U. Rasthoffer, F. Xiao, G. Winckelmans and the anonymous reviewers, whose inquiries, suggestions and ideas have helped to make this paper better.

This paper is dedicated to the memory of Jan Stępień (1937-2016).

\section*{List of Acronyms}

\begin{itemize}
\item ADM - Approximate Deconvolution Method
\item LS - Level Set (interface tracking method)
\item VOF - Volume of Fluid (interface tracking method)
\item CLSVOF - Coupled Level Set Volume of Fluid 
\item LES - Large Eddy Simulation
\item DNS - Direct Numerical Simulation
\item GFM - Ghost Fluid Method
\item MGCG - Multigrid Conjugate Gradients
\item WENO - Weighted Essentialy Non-Oscillatory
\item MAC - Marker-and-cell
\item SGSD - Sub-grid Surface Dynamics
\item SGS - Sub-grid scale
\end{itemize}

\bibliography{tau} %was aniszewski, it's supposed to be a bbl file (??)
\end{document}